\documentclass[aps,prd,superscriptaddress,twocolumn,preprintnumbers,nofootinbib,amsmath,amssymb,10pt]{revtex4-2}
\usepackage{graphicx}
\usepackage{xcolor}
\usepackage[normalem]{ulem}
\usepackage{hyperref}
\hypersetup{pdftitle={Localization in the Dirac spectrum and gauge-field topology}}

\usepackage{mathrsfs,multirow}  
\newcommand{\f}[2]{\frac{#1}{#2}}

\newcommand{\la}{\langle}

\newcommand{\ra}{\rangle}

\renewcommand{\Re}{{\rm Re}\,}

\newcommand{\tr}{{\rm tr}\,}

\usepackage{epstopdf}
\usepackage{placeins}
\usepackage{multirow}
\usepackage{makecell}
\usepackage{cleveref}
\usepackage{xcolor}

\newcommand{\beq}{\begin{eqnarray}}
\newcommand{\eeq}{\end{eqnarray}}
\newcommand{\beqnn}{\begin{eqnarray*}}
\newcommand{\eeqnn}{\end{eqnarray*}}

\renewcommand{\L}{\scriptscriptstyle{\rm L}}
\newcommand{\I}{\scriptscriptstyle{\rm I}}
\renewcommand{\c}{\scriptscriptstyle{\rm c}}
\newcommand{\Q}{\scriptscriptstyle{\rm Q}}
\newcommand{\PL}{P}
\newcommand{\PE}{L}

\newcommand{\tcz}{T_{\c}(0)}

\begin{document}
\title{Localization in the Dirac spectrum and gauge-field topology}

\author{Claudio Bonanno}
\email{claudio.bonanno@csic.es}
\affiliation{Instituto de F\'isica Te\'orica UAM-CSIC, c/ Nicol\'as
  Cabrera 13-15, Universidad Aut\'onoma de Madrid, Cantoblanco,
  E-28049 Madrid, Spain}

\author{Matteo Giordano}
\email{giordano@bodri.elte.hu}
\affiliation{Institute of Physics and Astronomy, ELTE E\"otv\"os
Lor\'and University, P\'azm\'any P\'eter s\'et\'any 1/A, H-1117,
Budapest, Hungary}

\date{\today}
\begin{abstract}
  We study localization of the low Dirac modes in 3+1 dimensional pure
  $\mathrm{SU}(3)$ gauge theory at zero and nonzero imaginary $\theta$
  angle, with the aim of better characterizing the relation between
  low-mode localization and topological features of gauge theories.
  We show that the mobility edge observed in the deconfined phase at
  $\theta=0$ is present also at nonzero $\theta$, appearing exactly at
  the deconfinement transition.  We find that the mobility edge is
  affected by topology only indirectly, through its effects on the
  ordering of the Polyakov loop.  Moreover, the change in the mobility
  edge is strongly correlated with the change in the Polyakov-loop
  expectation value, both as one moves along the critical line, and as
  one departs from it toward higher temperatures. This further
  strengthens the connection between low-mode localization and
  deconfinement, showing in particular the key role played by the
  ordering of the Polyakov loop.
\end{abstract}

\maketitle

\section{Introduction}
\label{sec:intro}

The discovery of a close relationship between the confining properties
and the localization properties of the low Dirac modes in a gauge
theory~\cite{Gockeler:2001hr,Gattringer:2001ia,GarciaGarcia:2005vj,
  GarciaGarcia:2006gr,Gavai:2008xe,Kovacs:2009zj,Kovacs:2010wx,
  Kovacs:2012zq,Giordano:2013taa,Nishigaki:2013uya,Ujfalusi:2015nha,
  Cossu:2016scb,Giordano:2016nuu,Kovacs:2017uiz,Holicki:2018sms,
  Giordano:2019pvc,Vig:2020pgq,Bonati:2020lal,Baranka:2021san,
  Kovacs:2021fwq,Cardinali:2021fpu,Baranka:2022dib,Kehr:2023wrs,
  Baranka:2023ani,Bonanno:2023mzj,Baranka:2024cuf} (see
Ref.~\cite{Giordano:2021qav} for a review) has given a new angle on
the finite-temperature transition and on the relation between
deconfinement and chiral symmetry restoration.  While delocalized in
the confined phase, in the deconfined phase of a gauge theory (or more
generally in a phase where the Polyakov loop gets ordered) the low
Dirac modes are localized, up to a ``mobility edge'', $\lambda_{\c}$,
in the spectrum. This moves toward zero as one approches the
(pseudo)critical temperature, $T_{\c}$, and vanishes (sometimes
abruptly~\cite{Bonati:2020lal,Kovacs:2021fwq}) as one crosses over to
the confined phase. This happens exactly at the critical temperature
when the transition is a genuine thermodynamic transition.\footnote{An
  exception is pure $\mathbb{Z}_2$ gauge theory in 2+1 dimensions,
  where $\lambda_{\c}$ increases as $T_c$ is approached from above, and
  disappears abruptly at $T_c$~\cite{Baranka:2024cuf}.} This suggests
that the localization centers for the low Dirac modes are intimately
related to the gauge-field structures responsible for confinement and
its loss, as it has been recently demonstrated in the simplest case of
$\mathbb{Z}_2$ gauge theory in 2+1
dimensions~\cite{Baranka:2024cuf}. The connection between
deconfinement and localization could then help in understanding the
microscopic mechanism(s) behind confinement in QCD and other gauge
theories. Given the central role of low Dirac modes in determining the
fate of chiral symmetry, it could also help in clarifying the relation
between chiral symmetry restoration and deconfinement.

Another aspect of gauge theories that is strongly affected by the
finite-temperature transition in QCD and in other gauge theories with
gauge group $\mathrm{SU}(N_c)$ are the topological features of typical
gauge configurations. It is well known that the magnitude and the
temperature dependence of the topological susceptibility change
dramatically across the transition, both in QCD and in pure
$\mathrm{SU}(N_c)$ gauge theory, with a strong suppression of
nontrivial topology above $T_{\c}$, getting stronger as the
temperature increases~\cite{DelDebbio:2004vxo,Lucini:2004yh,
  Bonati:2013tt,Bonati:2015vqz,Petreczky:2016vrs,Frison:2016vuc,
  Borsanyi:2016ksw,Bonati:2018blm,Lombardo:2020bvn,Borsanyi:2021gqg,
  Athenodorou:2022aay,Borsanyi:2022fub,Bonanno:2023hhp,Bonanno:2024zyn,
  Kotov:2025ilm}. Moreover, an ideal instanton gas-like behavior
emerges sufficiently far above $T_{\c}$~\cite{Bonati:2013tt,
  Bonati:2015vqz,Bonati:2018blm,Athenodorou:2022aay,Borsanyi:2022fub},
although with properties different from those of the semiclassical
dilute instanton gas~\cite{Gross:1980br, Boccaletti:2020mxu}. This is
actually expected in QCD because of the (approximate) restoration of
chiral symmetry in the high-temperature phase~\cite{Kanazawa:2014cua,
  Giordano:2024jnc}.

A third aspect affected by the transition is the behavior of the
spectral density of the Dirac operator near zero, that may be related
to the change in both the localization properties of Dirac modes and
the gauge-field topology. Besides a general depletion of the low-mode
region, expected in connection with chiral symmetry restoration, above
$T_{\c}$ the near-zero region displays a singular, power-law divergent
spectral peak, both in pure
gauge~\cite{Edwards:1999zm,Alexandru:2015fxa,Kovacs:2017uiz,
  Alexandru:2019gdm,Alexandru:2021pap,
  Alexandru:2021xoi,Vig:2021oyt,Kovacs:2021fwq} and in the presence of
dynamical fermions~\cite{Cossu:2013uua,Dick:2015twa,
  Tomiya:2016jwr,Aoki:2020noz,Ding:2020xlj,Kaczmarek:2021ser,
  Meng:2023nxf,Kaczmarek:2023bxb,Alexandru:2024tel}. It has been
suggested that this peak originates from fluctuations of unit charge
in the gauge-field topology and the associated (localized) Dirac
zero-modes~\cite{Edwards:1999zm,Kovacs:2017uiz,Vig:2021oyt,Kovacs:2023vzi}
(see Refs.~\cite{Alexandru:2015fxa,Alexandru:2021pap,
  Alexandru:2021xoi,Alexandru:2019gdm,Meng:2023nxf} for an alternative
proposal). Moreover, numerical results indicate that the localization
properties of the peak modes are
nontrivial~\cite{Alexandru:2021pap,Alexandru:2021xoi,Meng:2023nxf},
with the lowest ones delocalized, or possibly with nontrivial
localization properties, and localized higher peak modes. It has been
proposed that a mobility edge near zero (but not at zero) separates
delocalized peak modes in the immediate vicinity of zero, and
localized modes higher up in the spectrum (of course, still below the
mobility edge $\lambda_{\c}$ in the bulk), if $\mathrm{U}(1)_A$
remains effectively broken in the chiral limit in the symmetric
phase~\cite{Giordano:2024jnc}.

Deconfinement, chiral symmetry restoration, low-mode localization, and
change in topology are all closely connected phenomena, and
understanding how they affect each other would help in figuring out
what happens at the transition at the microscopic level, and possibly
in identifying one of them as being responsible for the others (if
there is such a thing). As a first step in this program, in this paper
we aim at investigating how the topological features of gauge fields
affect the mobility edge $\lambda_{\c}$, separating localized low
modes from delocalized bulk modes. For simplicity, we do this in the
``quenched'' limit of pure $\mathrm{SU}(3)$ gauge theory, probed with
external staggered fermions.  The strategy is to artificially increase
the instanton content of gauge configurations by switching on an
imaginary $\theta$-term, and check how $\lambda_{\c}$ responds.

Gauge theories and similar systems in the presence of a topological
term at imaginary $\theta$-angle, $\theta_{\I}$, have been intensely
studied on the lattice~\cite{Bhanot:1984rx,Azcoiti:2002vk,
  Imachi:2006qq,Azcoiti:2007cg,Alles:2007br,Aoki:2008gv,Vicari:2008jw,
  Panagopoulos:2011rb,Azcoiti:2012ws,DElia:2012pvq,Bonati:2013tt,
  DElia:2013uaf,Alles:2014tta,Bonati:2015sqt,Bonati:2016tvi,
  Bonanno:2020hht,Bonanno:2023hhp,Bonanno:2024ggk,Hirasawa:2024fjt},
mainly in relation with theoretical and phenomenological issues
connected to the dependence on a real $\theta$-angle, such as the
$``\mathrm{U}(1)_A$ problem'' or the ``strong-$CP$ problem''. Indeed,
this strategy allows one, by means of analytic continuation, to obtain
information on the real-$\theta$ case, that cannot be directly
simulated due to the ``complex-action problem'', or ``sign problem''.

While in this work we are mainly using the topological term to control
the amount of topological excitations in the system, it is worth
summarizing the relevant results concerning thermodynamics in its
presence. Pure $\mathrm{SU}(3)$ gauge theory at nonzero $\theta_{\I}$
displays the same kind of first-order transition observed at
$\theta_{\I}=0$~\cite{Boyd:1996bx}, with spontaneous breaking of
center symmetry due to ordering of the Polyakov loop. The expectation
value of this observable changes discontinuously at the critical
temperature, from zero in the low-temperature, confined phase to a
nonzero value in the high-temperature, deconfined phase. The critical
temperature, $T_{\c}(\theta_{\I})$, increases with $\theta_{\I}$, as a
larger imaginary $\theta$-angle makes it more favorable for the system
to remain in the confined phase~\cite{DElia:2012pvq,DElia:2013uaf}.
Accordingly, the magnitude of the Polyakov loop at fixed temperature
in the deconfined phase decreases with
$\theta_{\I}$~\cite{DElia:2012pvq, DElia:2013uaf}, as the increase in
topological content tends to drive the system back to the confined
phase. The phase diagram at physical, real $\theta$-angle can be
reconstructed to some extent by means of analytic
continuation~\cite{DElia:2012pvq,DElia:2013uaf,Bonanno:2020hht,
  Hirasawa:2024fjt}, and more directly by using reweighting
methods~\cite{DElia:2013uaf,Otake:2022bcq} or the subvolume
method~\cite{Kitano:2021jho,Yamada:2024vsk,Yamada:2024pjy}. This
interesting but difficult issue is outside of the scope of the present
paper. Here we are interested simply in increasing the nontrivial
topological content of gauge configurations, which is achieved in a
controllable way through an imaginary $\theta$-angle.

In the absence of a topological term, the spectrum of the staggered
operator in pure $\mathrm{SU}(3)$ gauge theory and the localization
properties of its eigenmodes have been studied in
Ref.~\cite{Kovacs:2017uiz}. A similar study for the overlap operator
was carried out in Ref.~\cite{Vig:2020pgq}. In both cases a mobility
edge was found in the low-lying spectrum, that moves toward the origin
as the temperature decreases. Extrapolating a fit to numerical data,
$\lambda_{\c}$ was seen to vanish at a temperature compatible with the
critical temperature in both cases. A more detailed study using the
overlap operator was carried out exactly at $T_{\c}$ by analyzing
separately the configurations belonging to the confined phase and to
the real and complex sectors of the deconfined
phase~\cite{Kovacs:2021fwq}. This showed that the mobility edge in the
real sector of the deconfined phase is nonzero at $T_{\c}$ (while it
is absent in the complex sectors), and disappears abruptly when
crossing over to the confined phase.

The observed universal nature of the connection between deconfinement
and low-mode localization leads one to expect that a mobility edge is
present in the staggered spectrum in the deconfined phase also in the
presence of a topological term in the action. The universality of this
connection is qualitatively explained by the ``sea/islands''
picture~\cite{Bruckmann:2011cc,Giordano:2015vla,Giordano:2016cjs,
  Giordano:2016vhx,Giordano:2021qav,Baranka:2022dib}. According to
this picture, low-mode localization in the deconfined phase results
from two effects: (1.)\ the opening of a pseudogap (i.e., a region of
low spectral density) in the Dirac spectrum as a consequence of the
ordering of the Polyakov loop; and (2.)\ the presence of localized
``islands'' of gauge-field fluctuations in the ``sea'' of ordered
Polyakov loops that are favorable for low modes.

From this point of view, the deconfined phase of $\mathrm{SU}(3)$
gauge theory at nonzero imaginary $\theta$ is no different than any
other deconfined phase, and low-mode localization should take
place. We expect then that a mobility edge appears in the real
Polyakov-loop sector at temperatures on the critical line,
$T_{\c}(\theta_{\I})$; and that it moves up in the spectrum as the
temperature increases, i.e., as the distance from the critical line
increases. If this is actually the case, it is an interesting question
whether a suitable definition of this distance would fully capture the
dependence of the mobility edge on thermal effects. In this context,
the most natural choice is the reduced temperature
$t(T,\theta_{\I})\equiv [T -
T_{\c}(\theta_{\I})]/T_{\c}(\theta_{\I})$.

Determining the $\theta$ dependence of $\lambda_{\c}$ does not present
any particular difficulty in itself. On the other hand, actually
understanding how the $\theta$-term affects $\lambda_{\c}$ seems
hopeless, since its introduction affects several features at the same
time: (1.)\ the critical temperature of the system, that increases
with $\theta_{\I}$; (2.)\ the ordering of the Polyakov loop, that
plays an essential role in low-mode localization; and (3.)\ the amount
of local thermal gauge-field fluctuations as well as of local
topological fluctuations, that is likely to affect the amount of
favorable localization centers.

Luckily enough, the answer turns out to be quite simple. As we will
demonstrate by means of numerical simulations, the dependence of
$\lambda_{\c} $ on $\theta_{\I}$ and on the temperature $T$ decomposes
into the sum of two terms,
$\lambda_{\c}|_{T,\,\theta_{\I}} \approx
\lambda_{\c}|_{T=T_{\c}(\theta_{\I}),\,\theta_{\I}} +
F(t(T,\theta_{\I}))$. The first term is the position of the mobility
edge at the $\theta_{\I}$-dependent critical temperature,
$T_{\c}(\theta_{\I})$. This is obtained following the approach used in
Ref.~\cite{Kovacs:2021fwq} at $\theta_{\I}=0$, by separating confining
gauge configurations and non-confining gauge configurations in the
real Polyakov-loop sector. As in that case, a mobility edge is found
only in the non-confining configurations. The second term is a
function of $T$ and $\theta_{\I}$ that depends only on the reduced
temperature $t$, and vanishes at $t=0$. This shows how indeed the
reduced temperature is the appropriate variable to describe the
influence of thermal fluctuations on the mobility edge.  Moreover, we
found that the two terms are strongly (directly) correlated with the
corresponding change in the Polyakov-loop expectation value, from
$\tcz$ to $T_{\c}(\theta_{\I})$ along the critical line for the first
term, and from the critical line ($t=0)$ along the temperature axis
($t> 0$) for the second term. This indicates that the increase with
$\theta_{\I}$ in local topology at a given temperature, $T$, affects
$\lambda_{\c}$ only indirectly, through its (dis)ordering effect on
the Polyakov loop configuration.

The plan of the paper is as follows. Sections~\ref{sec:SU3} and
\ref{sec:stag} are devoted to introductory material, with a brief
description of pure $\mathrm{SU}(3)$ gauge theory with an imaginary
$\theta$-term in Sec.~\ref{sec:SU3}, and of localization of staggered
eigenmodes in Sec.~\ref{sec:stag}. In Sec.~\ref{sec:num} we present
our numerical results, that we discuss in Sec.~\ref{sec:disc}.
Finally, in Sec.~\ref{sec:concl} we draw our conclusions and discuss
future directions for our studies.

\section{Lattice \texorpdfstring{$\mathrm{SU}(3)$}{SU(3)} gauge
  theory with a \texorpdfstring{$\theta$}{theta}-term}
\label{sec:SU3}

We discretize $\mathrm{SU}(3)$ gauge theory with a topological term on
a hypercubic, $N_s^3\times N_t$ lattice. Sites are denoted as
$n=(\vec{x},t)$, with $0\le x_i\le N_s-1$ for $i=1,2,3$, and
$0\le t\le N_t-1$.  Link variables $U_\mu(n)\in\mathrm{SU}(3)$,
$\mu=1,\ldots, 4$, are assigned to the oriented lattice edges
connecting sites $n$ and $n+\hat{\mu}$, where $\hat{\mu}$ are the unit
vectors parallel to the lattice axes;
$U_{-\mu}(n)=U_\mu(n-\hat{\mu})^\dag$ are assigned to the oppositely
oriented lattice edges. Periodic boundary conditions are imposed in all
directions. The action is
\begin{equation}
  \label{eq:tot_action}
  S = S_{\mathrm{W}} - i \tilde{\theta}_{\L} Q_{\L}\,, 
\end{equation}
where $S_{\rm W}$ is the Wilson discretization of the Yang--Mills action,
\begin{equation}
  \label{eq:SWilson}
  S_{\mathrm{W}} = \beta\sum_n
  \sum_{1\le\mu<\nu\le 4}        \left(1-\f{1}{3}\Re\tr U_{\mu\nu}(n)\right)\,,
\end{equation}
where $\beta$ is the lattice coupling, the sums are over
all lattice sites and over pairs of (positive) directions,
respectively, and $U_{\mu\nu}(n)$ is the usual plaquette variable,
\begin{equation}
  \label{eq:plaq}
  U_{\mu\nu}(n) = U_\mu(n)U_\nu(n+\hat{\mu})U_{-\mu}(n+\hat{\mu}+\hat{\nu}) U_{-\nu}(n+\hat{\nu})\,;
\end{equation}
and $Q$ is the topological charge, defined here, as in
Refs.~\cite{Panagopoulos:2011rb,DElia:2012pvq,DElia:2013uaf}, as the
clover-discretized version of the field-theoretic continuum
expression, i.e., $Q_{\L} = \sum_n q_{\L}(n)$ with
\begin{equation}
  \label{eq:top_charge}
  q_{\L}(n) = 
  -\f{1}{2^9 \pi^2}  \sum_{\mu,\nu,\rho,\sigma=\pm 1}^{\pm 4}
  \tilde{\varepsilon}_{\mu\nu\rho\sigma}
  \mathrm{tr}\left[U_{\mu\nu}(n)U_{\rho\sigma}(n)\right]\,,  
\end{equation}
where
$\tilde{\varepsilon}_{\mu\nu\rho\sigma}=\varepsilon_{\mu\nu\rho\sigma}$
is the usual totally antisymmetric Levi-Civita tensor if all indices
are positive, and
$\tilde{\varepsilon}_{(-\mu)\nu\rho\sigma}=-\tilde{\varepsilon}_{\mu\nu\rho\sigma}$.
Expectation values in the model defined by Eq.~\eqref{eq:tot_action}
are denoted with $\la\ldots\ra$.

The lattice operator $q_{\L}(n)$ is related to the topological charge
density operator $q(x)$ in the continuum as
$q_{\L}(n) \sim a^4 Z_{\Q}(\beta)q(an) +
O(a^6)$~\cite{Campostrini:1988cy}, where $a$ is the lattice spacing
and $Z_{\Q}(\beta)$ a suitable renormalization constant that tends to
1 in the continuum limit, $\lim_{\beta\to\infty} Z_{\Q}(\beta)=1$. The
physical topological charge $Q$ is defined as the integer closest to
$Q_{\L}$ measured after cooling~\cite{Ilgenfritz:1985dz,Teper:1985gi},
and the renormalization constant $Z_{\Q}$ is obtained as
$ Z_{\Q} = \la Q Q_{\L}\ra/\la Q^2\ra$~\cite{Panagopoulos:2011rb},
where the expectation values are computed at $\tilde{\theta}_{\L}=0$.
Values of $Z_{\Q}$ for the range of lattice couplings considered in
this paper can be found in Ref.~\cite{DElia:2012pvq}. We work here at
imaginary $\tilde{\theta}_{\L}=-i\theta_{\L}$. Accounting for the
renormalization of the topological charge, the relevant physical
imaginary angle at a given lattice spacing is related to $\theta_{\L}$
and $\beta$ as $\theta_{\I}= Z_{\Q}(\beta) \theta_{\L}$.

The $\mathrm{SU}(3)$ gauge theory with an imaginary $\theta$-term
displays a first-order transition at a critical lattice coupling
$\beta_{\c}$, dependent on $N_t$ and $\theta_{\I}$.  At
$\beta<\beta_{\c}$ the system is in a confined phase with unbroken
center symmetry, i.e., it is invariant under the center transformation
\begin{equation}
  \label{eq:center}
 U_4(\vec{x},N_t-1)\to z U_4(\vec{x},N_t-1)\,,\quad \forall\vec{x} \,,
\end{equation}
where $z\in\mathbb{Z}_3$ is an element of the gauge group center. In
this phase, the Polyakov loop,
\begin{equation}
  \label{eq:PL}
  P(\vec{x}) \equiv \f{1}{3}\tr\left(\prod_{t=0}^{N_t-1}U_4(\vec{x},t)\right)\,,
\end{equation}
has vanishing expectation value. At $\beta>\beta_{\c}$ center symmetry
breaks down spontaneously, due to the Polyakov loops $P(\vec{x})$
ordering along one of the center elements
$\{1,e^{\pm i\f{2\pi}{3}}\}$.\footnote{For a study of the relation
  between deconfinement and ordering of the Polyakov loop in
  higher-order representations in pure gauge theory and at the physical
  point see Refs.~\cite{Gupta:2007ax,Mykkanen:2012ri,
    Hanada:2023rlk,Hanada:2023krw}.} The real sector with $P(\vec{x})$
favoring the value 1 is the sector that would be selected by heavy
fermions in the ``quenched'' limit of infinite mass, and for this
reason it is sometimes referred to as the ``physical'' sector.  The
critical coupling $\beta_{\c}(N_t,\theta_{\I})$ has been determined in
Refs.~\cite{DElia:2012pvq,DElia:2013uaf} for several $N_t$ and a range
of $\theta_{\I}$. At $\beta_{\c}$ the expectation value of the
magnitude of the Polyakov loop, $\la |\PL|\ra$, where
$\PL\equiv \f{1}{N_s^3}\sum_{\vec{x}}P(\vec{x})$, changes
discontinuously from zero to a nonzero, presumably
$\theta_{\I}$-dependent value. In the deconfined phase $\la |\PL|\ra$
decreases quadratically with $\theta_{\I}$ for small $\theta_{\I}$ at
fixed $T$~\cite{DElia:2012pvq,DElia:2013uaf}. This means that the
introduction of an imaginary $\theta$-term makes the Polyakov loop
less ordered.

\section{Localization of staggered eigenmodes}
\label{sec:stag}

The staggered discretization of the Dirac operator is
\begin{equation}
  \label{eq:stag}
  D = \f{1}{2}\sum_{\mu=1}^4\eta_\mu\left( U_\mu T_\mu - T_\mu^\dag U_\mu^\dag\right)\,,
\end{equation}
where $T_\mu $ is the translation operator in direction $\mu$, with
boundary conditions periodic in space and antiperiodic in time
understood; and $\eta_\mu$ are diagonal matrices, trivial in color
space, with entries the usual staggered phases
$\eta_\mu(n)=(-1)^{\sum_{\alpha<\mu} n_\alpha}$. The staggered
operator is anti-Hermitian, and obeys $\{D,\varepsilon\}=0$, where
$\varepsilon$ is a diagonal, color-trivial matrix with entries
$\varepsilon(n) = (-1)^{\sum_{\alpha} n_\alpha}$. Its eigenvalues are
therefore purely imaginary, and its spectrum is symmetric about
zero. In fact, if $\psi_l$ is an eigenvector of $D$ with eigenvalue
$ia\lambda_l$, then $\varepsilon\psi_l$ is an eigenvector with
eigenvalue $-ia\lambda_l$.

The localization properties of the staggered eigenmodes are determined
by the volume scaling of their ``size'', suitably averaged over gauge
configurations. A practical definition of the mode size is obtained
starting from the inverse participation
ratio~\cite{thouless1974electrons,lee1985disordered,kramer1993localization},
\begin{equation}
  \label{eq:IPR}
  \mathrm{IPR}_l \equiv a^{-4}\sum_n\Vert \psi_l(n) \Vert^4\,,
\end{equation}
where $\Vert \psi_l(n) \Vert^2= \sum_c |(\psi_l)_c(n)|^2$, with the
sum running over the color index, $c=1,2,3$, is the gauge-invariant
local magnitude of the eigenmode. The inverse of the IPR is a good
definition of the mode size: it is easy to check that if a mode has
magnitude uniformly distributed in a region of four-volume $V_0$, then
its IPR satisfies $(\mathrm{IPR})^{-1} = V_0$. The localization
properties of the modes in a given spectral region are determined by
the volume scaling of the average $\mathrm{IPR}$ computed locally in
the spectrum, $\overline{\mathrm{IPR}}(\lambda,L_s)$,\footnote{The
  dependence of the various expectation values on $T$ and
  $\theta_{\I}$ is left implicit. As the normalized spectral density
  is expected to depend mildly on the spatial volume, we leave its
  dependence on $L_s$ implicit, too.}
\begin{equation}
  \label{eq:size}
  \overline{\mathrm{IPR}}(\lambda,L_s) \equiv
  \f{1}{\f{V}{T}\rho(\lambda)}\left\la \sum_l\delta(\lambda-\lambda_l) \mathrm{IPR}_l\right\ra 
  \,,
\end{equation}
where $L_s=aN_s$, $V=L_s^3$ is the physical volume, and
$\rho(\lambda)$ is the spectral density normalized by the
four-volume, 
\begin{equation}
  \label{eq:sp_dens}
  \rho(\lambda) \equiv 
\f{T}{V}  \left\la \sum_l\delta(\lambda-\lambda_l)\right\ra    \,.
\end{equation}
For large $L_s$ one has
$\overline{\mathrm{IPR}}(\lambda,L_s)\sim L_s^{-D_2(\lambda)}$, where
$D_2$ is the fractal dimension of the modes. If $D_2(\lambda)=0$ the
mode size is volume independent, and the modes are localized; if
$D_2(\lambda)=3$ the modes spread out over the whole lattice and are
delocalized. For intermediate values the modes are known as
``critical'' in the condensed-matter literature: this is the kind of
behavior found at a mobility edge, i.e., a point in the spectrum
separating localized and delocalized modes, where the localization
length diverges and a phase transition (``Anderson transition'') is
encountered along the spectrum~\cite{Evers:2008zz}.

A practically more convenient way to detect localization is through
its connection with the (bulk) statistical properties of the
spectrum~\cite{altshuler1986repulsion}. In fact, localized modes
fluctuate independently of each other under local changes in the gauge
configuration, and so the corresponding eigenvalues are expected to
obey Poisson statistics. On the other hand, fluctuations of
delocalized modes under changes in the gauge configuration are
strongly correlated, and one expects the corresponding eigenvalues to
obey the same statistics as those of the appropriate ensemble of
random matrix theory (RMT)~\cite{mehta2004random,
  Verbaarschot:2000dy}, as determined by the symmetry class of the
system. In the case of the staggered operator in the background of
$\mathrm{SU}(3)$ gauge fields, these are the unitary class, and the
Gaussian Unitary Ensemble~\cite{Verbaarschot:2000dy}.

The statistical properties of the spectrum are most easily unveiled by
making use of the unfolding procedure, that reveals their universal
features. Unfolding is the monotonic mapping $\lambda_n\to x_n$ of the
spectrum defined by
\begin{equation}
  \label{eq:unfolding}
  x_n =    \f{V}{T}\int_{\lambda_{\mathrm{min}}}^{\lambda_n} d\lambda\,
 \rho(\lambda)  \,,
\end{equation}
that makes the density $r_u(x)$ of unfolded eigenvalues identically
equal to 1, i.e., $r_u=\f{V}{T}\rho\f{d\lambda}{dx} = 1$. In
particular, the probability distribution $p(s)$ of the unfolded
spacings, i.e., of the spacing between consecutive unfolded
eigenvalues, $s_i \equiv x_{i+1}-x_i$, is known for both Poisson and
RMT statistics. For Poisson statistics, $p_{\mathrm{P}}(s)=e^{-s}$;
for RMT statistics a closed form is not available, but a good
approximation is provided by the Wigner surmise
$p_{\mathrm{RMT}}(s) = c_1 s^{\beta_{\mathrm{D}}} e^{-c_2s^2}$,
with $\beta_{\mathrm{D}}=2$ for the unitary class, and $c_1$ and
$c_2$ determined by the normalization condition
$\int_0^\infty ds\,p(s)=1$, and by the fact that
$\int_0^\infty ds\,p(s)s=1/ r_u = 1$.

\begin{table*}[!htb]
  \small
  \begin{center}
    \begin{tabular}{|c|c|c|c|c|c|c|c|}
      \hline
      &&&&&&&\\[-1em]
      $N_t$ & $N_s$ & $\beta$ & $\theta_{\L}$ & $\theta_{\I} = Z_{\Q}(\beta) \theta_{\L}$ &   $t(T,\theta_{\I}) = T/T_{\c}(\theta_{\I})-1 $ &  $T_{\c}(\theta_{\I})/\tcz $  & $a\lambda_{\c}$ 
      \\
      \hline
      \hline
      \multirow{4}{*}{4} & \multirow{4}{*}{24}
        & 5.6911 & 0  & 0         & \multirow{4}{*}{0.00} & 1 & 0.202(16) \\
      && 5.7092 & 15 & 1.142(19) & & 1.0395(11) & 0.220(13) \\
      && 5.7248 & 20 & 1.566(23) & & 1.0746(10) & 0.235(14) \\
      && 5.7447 & 25 & 2.025(30) & & 1.1209(10) & 0.246(16) \\
      \hline
      \hline
      \multirow{4}{*}{4} & \multirow{4}{*}{24}
        & 5.7204 & 0  & 0         & \multirow{4}{*}{0.07} & 1 & 0.2900(21) \\
      && 5.7391 & 15 & 1.207(20) & & 1.0395(11) & 0.2961(34) \\
      && 5.7550 & 20 & 1.663(20) & & 1.0746(10) & 0.3038(18) \\
      && 5.7762 & 25 & 2.177(19) & & 1.1209(10) & 0.3129(20) \\
      \cline{1-5}
      \hline
      \hline
      \multirow{4}{*}{4} & \multirow{4}{*}{24}
        & 5.7407 & 0  & 0         & \multirow{4}{*}{0.12} & 1 & 0.3213(20) \\
      && 5.7597 & 15 & 1.260(14) & & 1.0395(11) & 0.3251(18) \\
      && 5.7762 & 20 & 1.742(15) & & 1.0746(10) & 0.3314(15) \\
      && 5.7970 & 25 & 2.282(18) & & 1.1209(10) & 0.3368(18) \\
      \hline
      \hline
      \multirow{4}{*}{4} & \multirow{4}{*}{24}
        & 5.7644 & 0  & 0         & \multirow{4}{*}{0.18} & 1 & 0.3438(21) \\
      && 5.7838 & 15 & 1.329(18) & & 1.0395(11) & 0.3525(25) \\
      && 5.8007 & 20 & 1.841(14) & & 1.0746(10) & 0.3567(17) \\
      && 5.8220 & 25 & 2.417(19) & & 1.1209(10) & 0.3635(19) \\
      \hline
    \end{tabular}
  \end{center}
  \caption{Summary of simulation parameters, and corresponding estimates
    of the mobility edge (see Sec.~\ref{sec:num_mobedge}).}
  \label{tab:summary_points}
\end{table*}

To check the localization properties of the modes one can pick a
feature of $p(s)$, e.g., its second moment, and monitor how its
average over configurations, computed locally in the spectrum, changes
as one moves along it. A numerically convenient choice in this
regard~\cite{Shklovskii:1993zz} is the integrated probability density,
$I_{s_0}$,
\begin{equation}
  \label{eq:IULSD}
  \begin{aligned}
    I_{s_0}(\lambda,L_s) &\equiv \int_0^{s_0} ds\,p(s;\lambda,L_s)\,,    \\
    p(s;\lambda,L_s) &\equiv \f{1}{\f{V}{T}\rho(\lambda)}
                       \left\la \sum_l \delta(\lambda-\lambda_l)\delta(s-s_l)\right\ra\,,
  \end{aligned}
\end{equation}
with $s_0\simeq 0.508$ chosen to maximize the difference between the
expectations for Poisson and (unitary) RMT statistics, i.e.,
$I_{s_0,\mathrm{P}}\simeq 0.398$ and
$I_{s_0,\mathrm{RMT}}\simeq 0.117$. As the system size increases,
$I_{s_0}(\lambda,L_s)$ tends to $I_{s_0,\mathrm{P}}$ or
$I_{s_0,\mathrm{RMT}}$, depending on the localization properties of
the modes in the spectral region under scrutiny. At a mobility edge,
$\lambda_{\c}$, $I_{s_0}(\lambda,L_s)$ is volume-independent and takes a
universal ``critical value'', $I_{s_0,c}$, that depends on the
symmetry class of the system; for the unitary class this has been
obtained in Ref.~\cite{Giordano:2013taa}. This allows one to determine
$\lambda_{\c}$ quite accurately and very efficiently as the point
where $I_{s_0}(\lambda_{\c},L_s)=I_{s_0,c}$~\cite{Kovacs:2017uiz}.

\section{Numerical results}
\label{sec:num}

We simulated pure $\mathrm{SU}(3)$ gauge theory at finite temperature
on $N_s^3\times N_t$ hypercubic lattices for temporal size $N_t=4$ and
spatial size $N_s=24$, for various values of the lattice coupling
$\beta$ and of the imaginary lattice $\theta$-parameter,
$\theta_{\L}$. The temperature $T=T(\beta,N_t)=1/[a(\beta)N_t]$ is
determined by the inverse temporal size and by the lattice coupling
$\beta$. For the scale setting we used the subpercent determinations
of $a(\beta)/r_0$ of Ref.~\cite{Necco:2001xg} (or interpolations
thereof), where $r_0$ is the Sommer scale~\cite{Sommer:1993ce}. We
chose for the bare angle the values $\theta_{\L}=0,15,20,25$, and
chose values of $\beta$ so that the dimensionless reduced temperature
\begin{equation}
  \label{eq:redT}
  t(T,\theta_{\I})=\f{T-T_{\c}(\theta_{\I})}{T_{\c}(\theta_{\I})}
\end{equation}
where $\theta_{\I}=Z_{\Q}(\beta)\theta_{\L}$, lied on trajectories of
constant $t=0,0.07,0.12,0.18$. For the renormalization constant
$Z_{\Q}(\beta)$, for the critical coupling $\beta_{\c}(\theta_{\I})$,
that determines the critical temperature
$T_{\c}(\theta_{\I})=1/[a(\beta_{\c}(\theta_{\I}))N_t]$, and for the
ratio $T/T_{\c}(\theta_{\I})$ we used the values found in
Refs.~\cite{DElia:2012pvq,DElia:2013uaf} (or interpolations
thereof). Our simulation parameters are summarized in
Tab.~\ref{tab:summary_points}.

For each choice of $(\beta,\theta_{\L})$ we then obtained the
low-lying spectrum of the staggered Dirac operator, computing its
lowest 200 positive eigenvalues (230 for $\theta_{\L}=25$) and
corresponding eigenvectors on 600 configurations, using the PARPACK
routine~\cite{maschho1996portable,PARPACK}.  Configurations were
separated by 100 update steps, one step consisting of 1
heat-bath~\cite{Creutz:1980zw,Kennedy:1985nu} and 4
over-relaxation~\cite{Creutz:1987xi} sweeps of the lattice. Both local
updates have been implemented \emph{\`a la}
Cabibbo--Marinari~\cite{Cabibbo:1982zn}, i.e., updating all the $3$
diagonal SU(2) subgroups of SU(3).

To unfold the spectrum we ranked all the eigenvalues of all the
configurations in the ensemble and replaced them by their rank divided
by the number of configurations~\cite{Kovacs:2012zq}. To check the
reliability of the method we monitored the average spacing computed
locally in the spectrum and verified that it was equal to 1 within
errors in the relevant spectral region where the mobility edge is
found. This also shows that the effects of taste symmetry, leading to
the formation of quartets of staggered eigenvalues, are negligible for
what concerns the determination of the mobility edge for our lattice
setup (see Ref.~\cite{Bonanno:2023mzj} for a detailed
discussion). These effects show up only at the lowest end of the
spectrum, where the average unfolded spacing deviates from 1.

\begin{figure*}[htb!]
  \centering
  \includegraphics[scale=0.45]{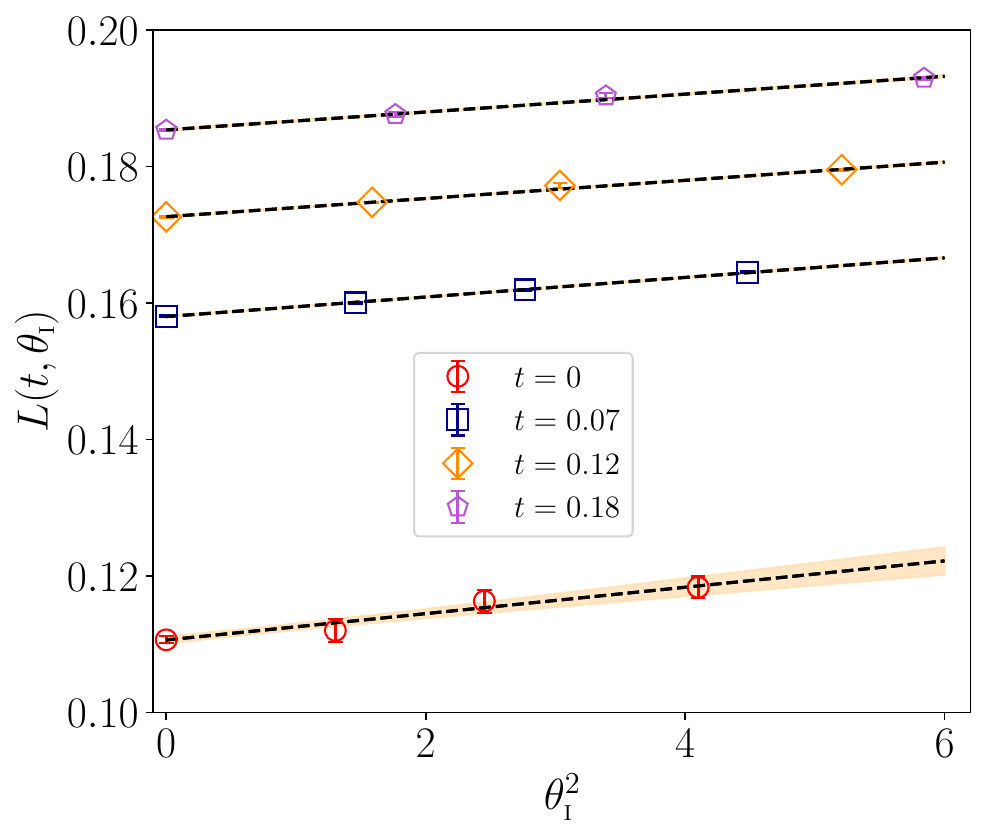}
  \hfil
  \includegraphics[scale=0.45]{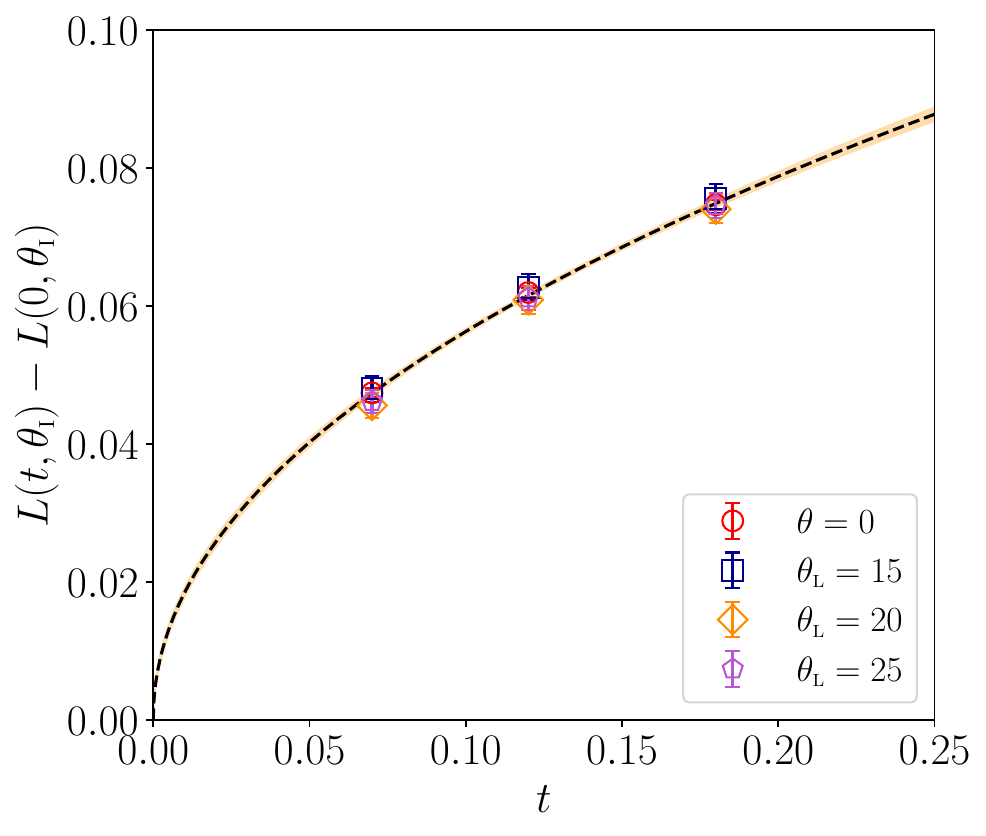}
  \caption{Left panel: Expectation value of the modulus of the
    spatially averaged Polyakov loop [see Eq.~\eqref{eq:PLev}] as a
    function of $\theta_{\I}^2$. Lines correspond to separate linear
    fits of data at constant $t$. Right panel: Expectation value of
    the modulus of the spatially averaged Polyakov loop for a given
    $\theta_{\I}$ minus its value at the corresponding critical
    temperature, $T_{\c}(\theta_{\I})$, as a function of the reduced
    temperature, $t$, for the available values of $\theta_{\I}$. The
    dashed line is a common fit with the power-law function $at^b$ to
    all the data. Our best fit yields $a = 0.172(5)$ and
    $b = 0.48(2)$, with a reduced chi-squared of $3.0/10.0$.}
  \label{fig:pol_th}
\end{figure*}

\subsection{Polyakov loop}
\label{sec:num_PL}

As a preliminary analysis, in the left panel of Fig.~\ref{fig:pol_th}
we show the expectation value of the magnitude of the spatially
averaged Polyakov loop,
\begin{equation}
  \label{eq:PLev}
  \PE(t,\theta_{\I})\equiv \left.\left\la
      |\PL|\right\ra \right|_{T=(1+t)T_{\c}(\theta_{\I}),\,\theta_{\I}}\,,
\end{equation}
as a function of $\theta_{\I}$ for the available values of $t$. This
shows that while this quantity decreases with $\theta_{\I}$ at fixed
temperature, $T$, as shown in Refs.~\cite{DElia:2012pvq,
  DElia:2013uaf}, it grows with $\theta_{\I}$ at fixed reduced
temperature, $t$, including along the critical line ($t=0$). The
dependence on $\theta_{\I}$ is clearly quadratic in the explored
range.  In the right panel of Fig.~\ref{fig:pol_th} we show $\PE$ as a
function of $t$ after subtracting its value at the critical
temperature, $\PE(t,\theta_{\I})-\PE(0,\theta_{\I})$. Since $Z_{\Q}$
depends on $\beta$ we do not have data at the same exact value of
$\theta_{\I}=Z_{\Q}\theta_{\L}$ for different temperatures, and so we
actually subtracted the value of $\PE$ at the critical temperature
corresponding to the same bare angle, $\theta_{\L}$. In spite of this,
the collapse of data points on a single, power-law curve is quite
clear, well within the numerical uncertainties. The reduced
temperature gives then a good measure of the ordering of the system
above $T_c$, independently of $\theta_{\I}$.

\begin{figure}[t!]
  \centering
  \includegraphics[scale=0.48]{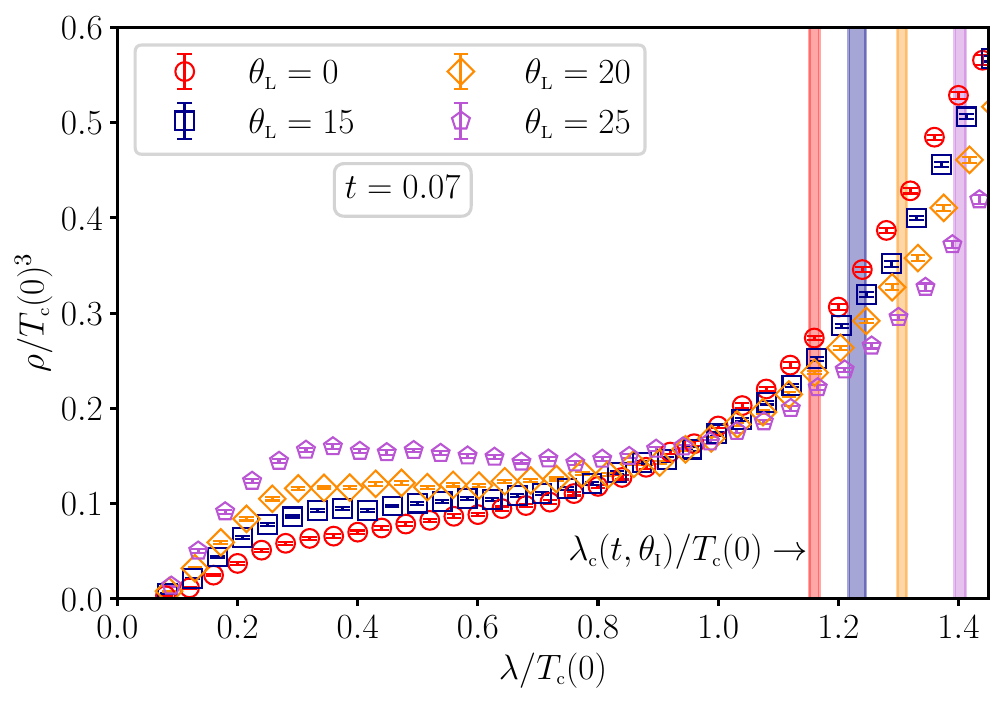}
  \caption{Spectral density at reduced temperature $t=0.07$ for all the
    available $\theta_{\L}$. The corresponding mobility edges and their
    uncertainties (see Sec.~\ref{sec:num_mobedge}) are shown by vertical
    bands of the same color as the data points.}
  \label{fig:spdens}
\end{figure}
\begin{figure}[t!]
  \centering
  \includegraphics[scale=0.45]{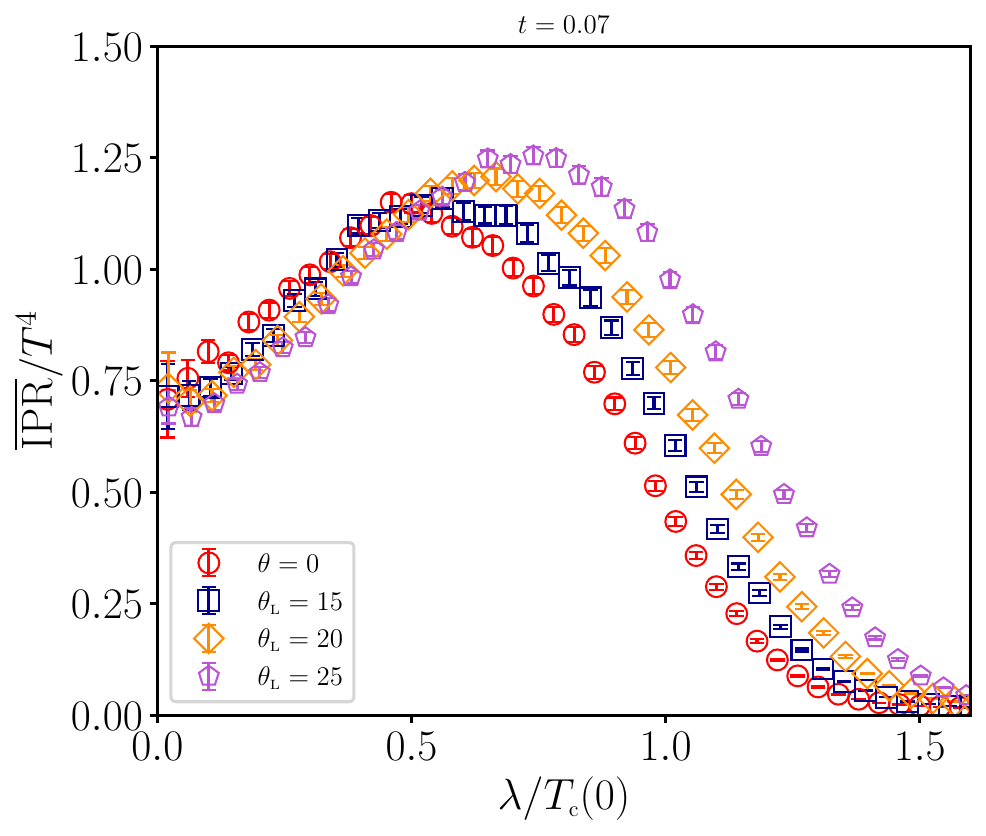}
  \caption{IPR at reduced temperature $t=0.07$ for all the available
    $\theta_{\L}$. }
  \label{fig:ipr}
\end{figure}

\begin{figure*}[bth!]
  \centering
  \includegraphics[scale=0.3]{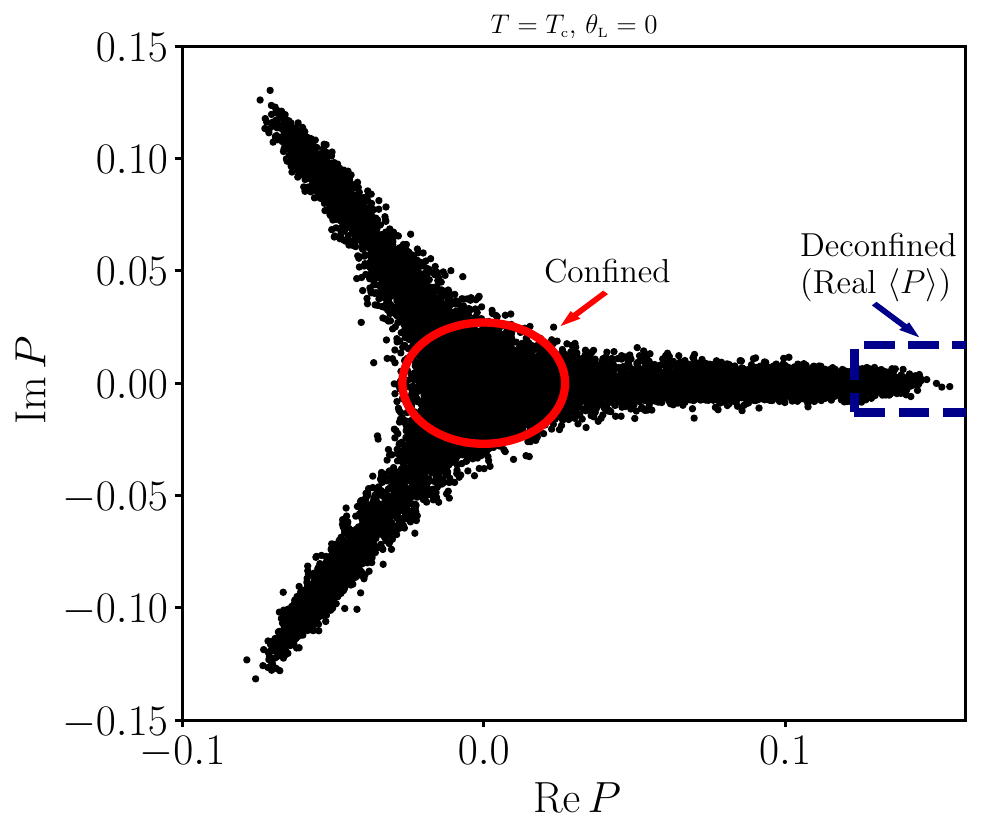}
  \includegraphics[scale=0.3]{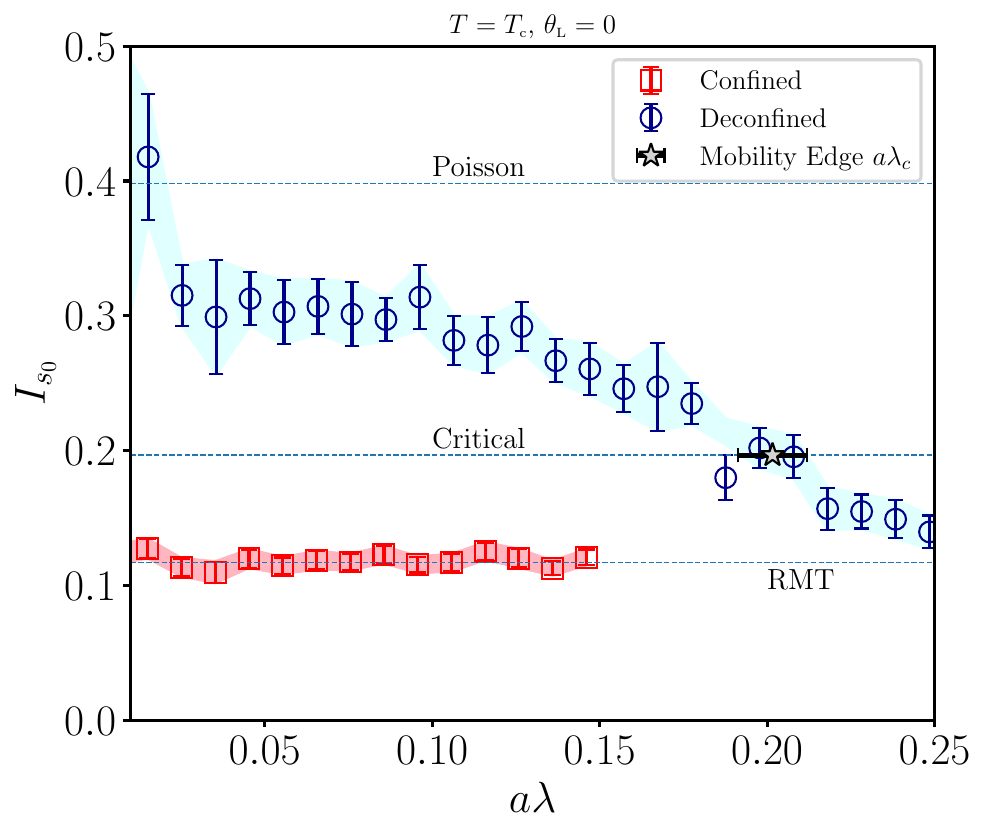}
  \includegraphics[scale=0.3]{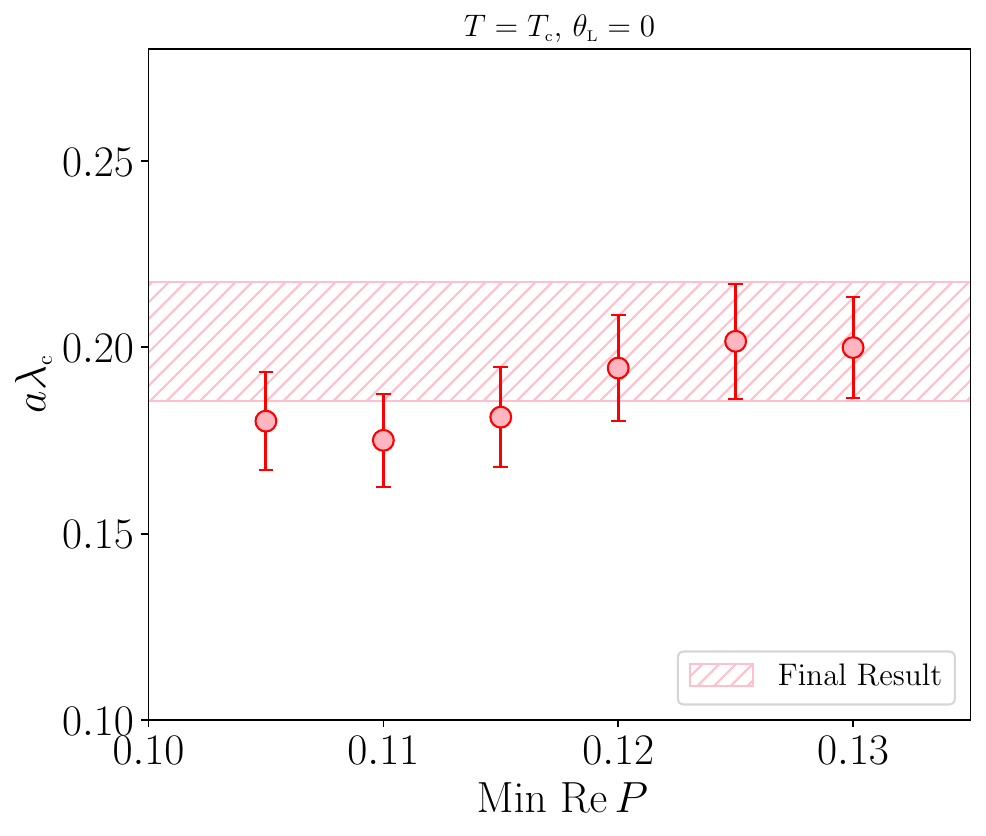}

  \includegraphics[scale=0.3]{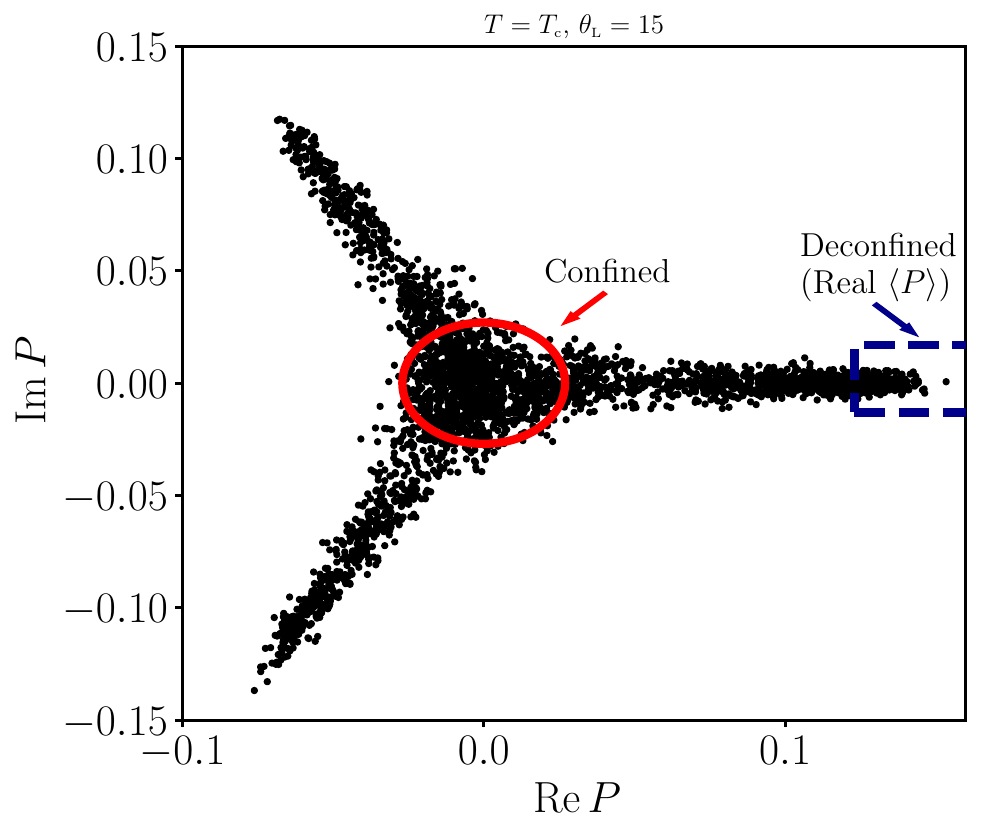}
  \includegraphics[scale=0.3]{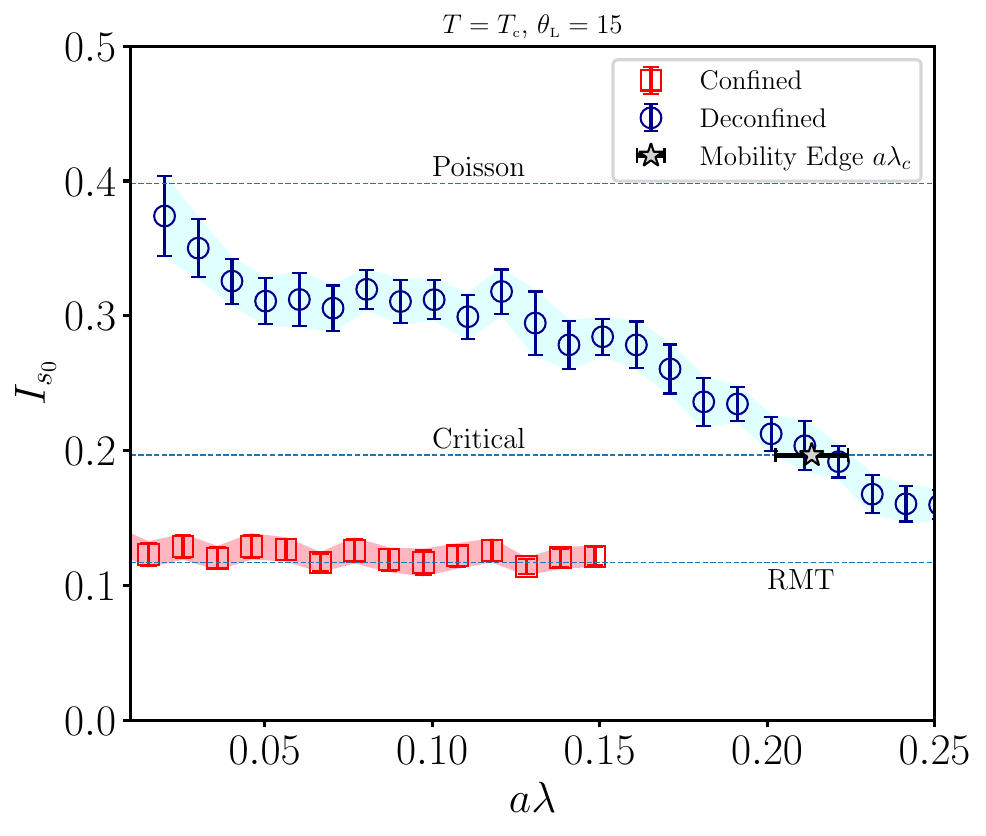}
  \includegraphics[scale=0.3]{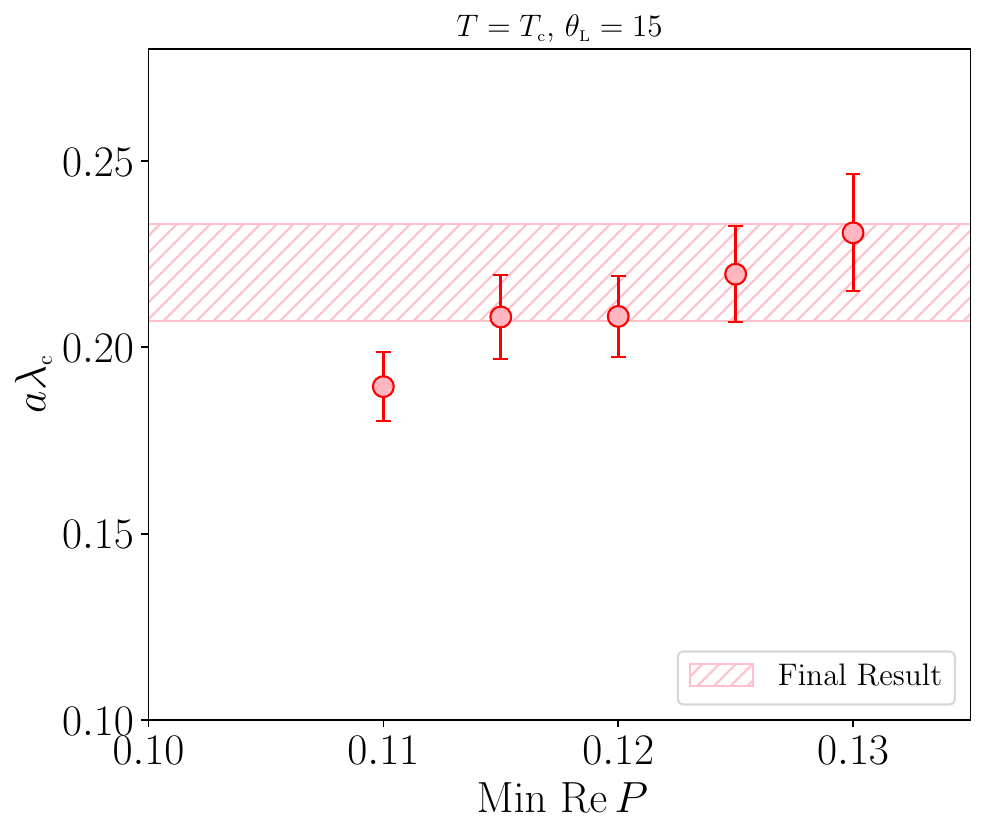}
  \caption{Appearance of the mobility edge at the critical temperature
    $T_{\c}(\theta_{\I})$ for $\theta_{\L}=0$ (top panels) and
    $\theta_{\L}=15$ (bottom panels). The left panels show how we
    separated configurations belonging to the confined phase from
    those belonging to the real sector of the deconfined phase. The
    center panels show the absence of a mobility edge in the confining
    configurations, and how we determined $\lambda_{\c}$ from
    $I_{s_0}$ for the deconfining ones for a specific choice of cutoff
    (here the center of the peak of the Polyakov-loop distribution at
    positive $\Re P$). The right panels show how our estimates of the
    mobility edge depend on the cutoff imposed on the Polyakov loop to
    assign a configuration to the deconfined phase.}
  \label{fig:loc_Tc_theta0}
\end{figure*}

\subsection{Spectral density}
\label{sec:num_spec}

In Fig.~\ref{fig:spdens} we show the spectral density $\rho$,
Eq.~\eqref{eq:sp_dens}, divided by $\tcz^3$ to make it dimensionless,
against the dimensionless ratio $\lambda/\tcz$. Here we use the fixed
scale $\tcz$ in order not to introduce additional dependences on $T$
and $\theta_{\I}$. The most interesting feature is the emergence of a
(rather broad) peak at the low end of the spectrum as one increases
$\theta_{\L}$. While nothing more than a barely visible shoulder at
$\theta_{\L}=0$, the structure is already quite clear at
$\theta_{\L}=20$. This structure is the counterpart for staggered
fermions on coarse lattices of the sharp near-zero peak observed with
overlap valence fermions in the background of pure $\mathrm{SU}(3)$
gauge configurations in Refs.~\cite{Edwards:1999zm,
  Alexandru:2015fxa,Alexandru:2019gdm,Alexandru:2021pap,Alexandru:2021xoi,
  Vig:2021oyt,Kovacs:2021fwq}, originating from the mixing of the zero
modes associated with weakly interacting topological
objects~\cite{Edwards:1999zm,Kovacs:2017uiz,Vig:2021oyt,Kovacs:2023vzi}. While
invisible on $N_t=4$ lattices at $\theta_{\L}=0$, it was shown in
Ref.~\cite{Kovacs:2017uiz} that the peak structure is already visible
at $N_t=6$ for $T=1.05\,\tcz$, becoming sharper as $N_t$ increases and
the lattice becomes finer at fixed $T$, with the staggered operator
becoming more accurate at resolving topological effects on the
spectrum. Here the emergence of the structure is instead driven by the
increase of $\theta_{\I}$ and so of the density of topological objects
in the system, further supporting the topological origin of the
near-zero peak.  Notice that the peak structure is well below the
mobility edge (determined in Sec.~\ref{sec:num_mobedge}).

\subsection{Mobility edge}
\label{sec:num_mobedge}

To demonstrate the presence of localized modes at the low end of the
spectrum, in Fig.~\ref{fig:ipr} we show the average IPR computed
locally in the spectrum, Eq.~\eqref{eq:size}, in units of $T^4$, for
$t=0.07$ and all available $\theta_{\L}$.  The normalization is chosen
in order to compare the size of the modes with the natural scale set
by the inverse temperature.  While, strictly speaking, the localized
nature of modes is revealed by how their size scales with the spatial
size of the system, it is clear that near-zero modes have a much
larger IPR and so are much smaller than bulk modes. Interestingly, the
IPR does not depend monotonically on $\lambda$, with a
$\theta_{\L}$-dependent maximum in the range
$\lambda/\tcz\sim 0.5 \,\text{--}\, 0.8$, that increases in height and
moves up in the spectrum as $\theta_{\L}$ increases. The smallest
modes become then smaller and are found farther away from zero, with a
spatial size $\ell$ of about
$\ell T \equiv \left(T^4/\overline{\mathrm{IPR}}\right)^{\f{1}{3}}
\sim 0.93\,\text{--}\, 0.96$, i.e., of the order of $1/T$. The lowest
modes have a smaller IPR and so a larger size
($\ell T \sim 1.1\,\text{--}\,1.13$), that depends mildly on
$\theta_{\L}$.  Such a mild $\theta_{\L}$-dependence is generally
displayed by modes in the region of the spectral peak seen in
Fig.~\ref{fig:spdens}.

\begin{figure*}[bth!]
  \centering
  \includegraphics[scale=0.49]{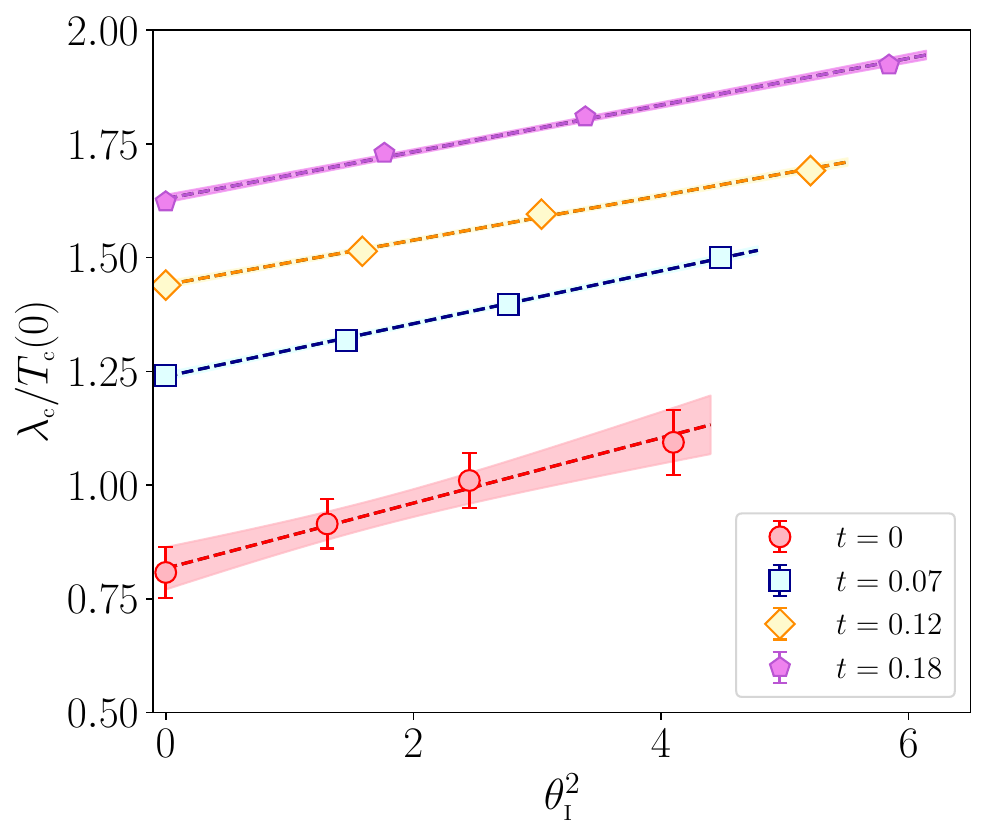}\hfil
  \includegraphics[scale=0.49]{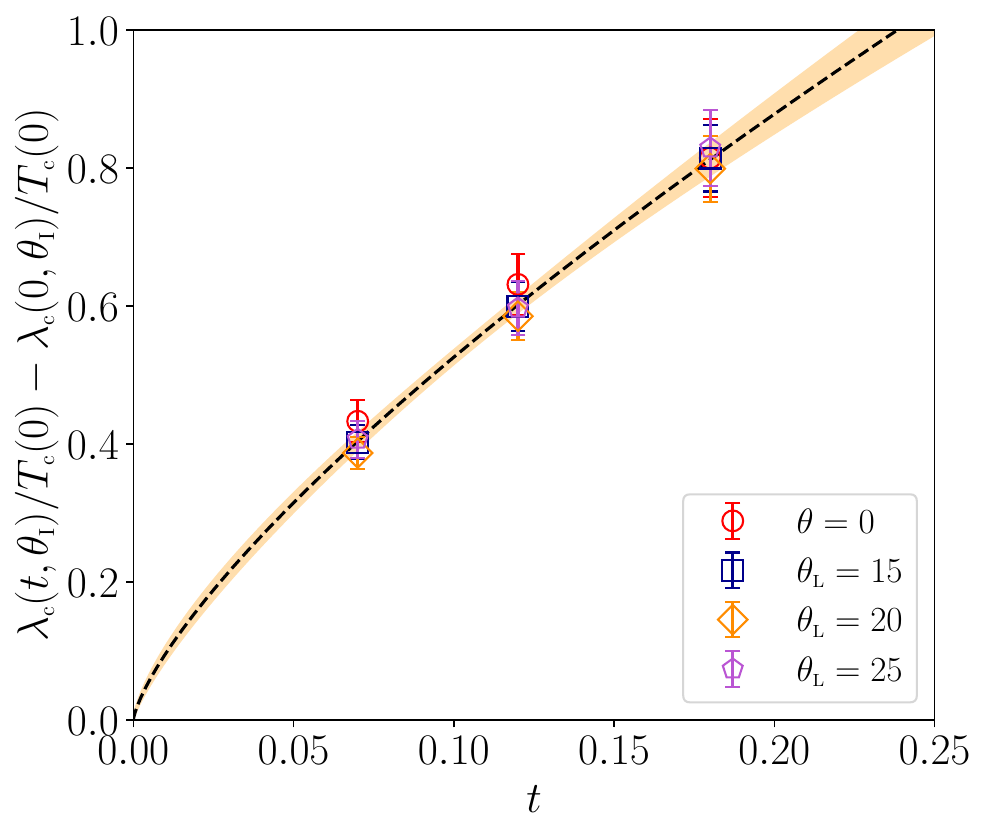}
  \caption{Left panel: Mobility edge in units of $T_{\c}(0)$ against
    $\theta_{\I}^2$. Lines correspond to separate linear fits of data
    at constant reduced temperature $t$. Right panel: Mobility edge in
    units of $T_{\c}(0)$ minus its value at $T_{\c}(\theta_{\I})$
    against the reduced temperature $t$. The curve corresponds to a
    fit of all the data points with the power-law function $a
    t^b$. Our best fit yields $a=2.89(30)$ and $b=0.74(5)$ with a
    reduced chi-squared of 2.4/10.}
  \label{fig:mod_indep}
\end{figure*}

We then proceeded with a systematic study of localization across the
available range of values of $\theta_{\I}$ and $t$, listed in
Tab.~\ref{tab:summary_points}. As a first step we studied localization
at $t=0$, i.e., at $T=T_{\c}(\theta_{\I})$. We followed the approach
of Ref.~\cite{Kovacs:2021fwq}, separating the non-confining gauge
configurations in the real Polyakov-loop sector from the confining
ones (see Fig.~\ref{fig:loc_Tc_theta0}, left panels), and identifying
$\lambda_{\c}$ as the crossing point of $I_{s_0}$ with its critical
value (see Fig.~\ref{fig:loc_Tc_theta0}, center panels).  For the
confining configurations $I_{s_0}=I_{s_0,\mathrm{RMT}}$ within errors
in the whole explored spectral range, indicating the absence of
localized low modes and of a mobility edge. In the non-confining
configurations, instead, for the low modes $I_{s_0}$ rises well above
$I_{s_0,\mathrm{RMT}}$ toward $I_{s_0,\mathrm{Poisson}}$, and crosses
the critical value, $I_{s_0,c}$, showing the presence of localized low
modes and of a mobility edge. The uncertainty on $\lambda_{\c}$ was
then estimated as the half-width of the interval between the crossing
points of the upper and lower ends of the error band of $I_{s_0}$ with
$I_{s_0,c}$. This whole procedure requires imposing a somewhat
arbitrary cutoff on the value of $\Re \PL$, near the peak in its
distribution corresponding to the real sector of the deconfined phase,
to safely label a certain configuration as non-confining, while
avoiding configurations intermediate between the confined and the
deconfined phase that play no role in the thermodynamic limit. To
check for systematic effects we have repeated the analysis changing
the value of the cutoff around the center of this peak (see
Fig.~\ref{fig:loc_Tc_theta0}, right panels). The results are within
the error band of $\lambda_{\c}$ obtained at the peak center, that we
then took with the corresponding error as our final estimate.

\begin{figure*}[tbh!]
  \centering
  \includegraphics[scale=0.49]{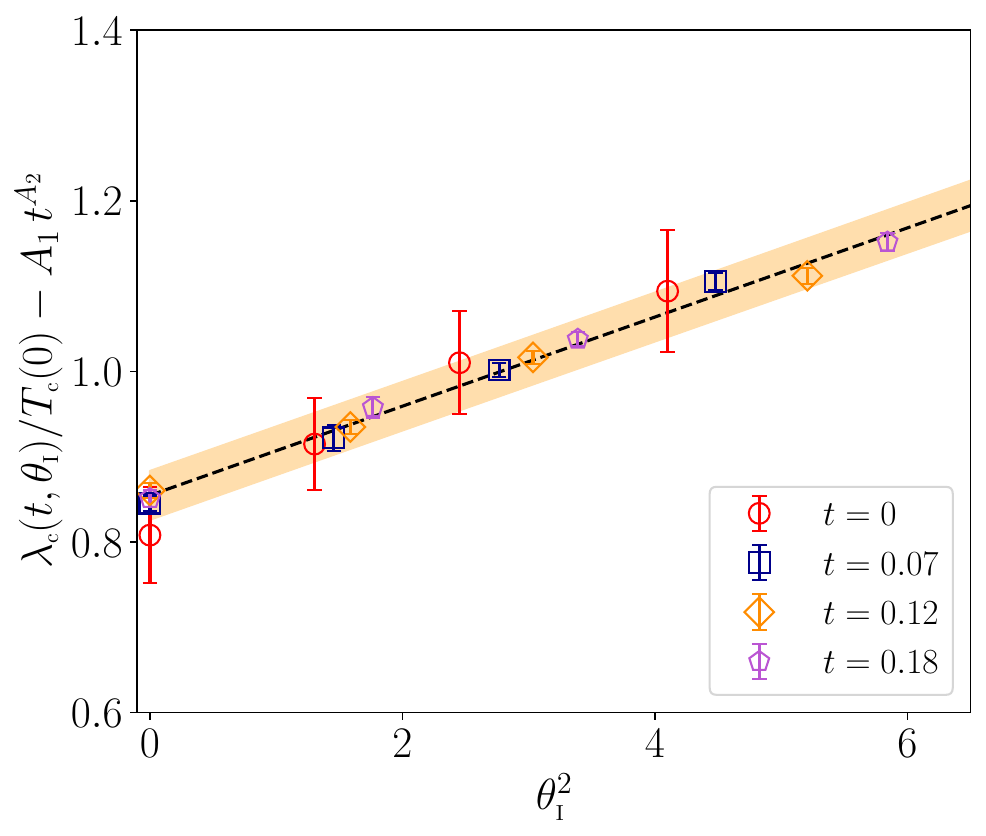}
  \includegraphics[scale=0.49]{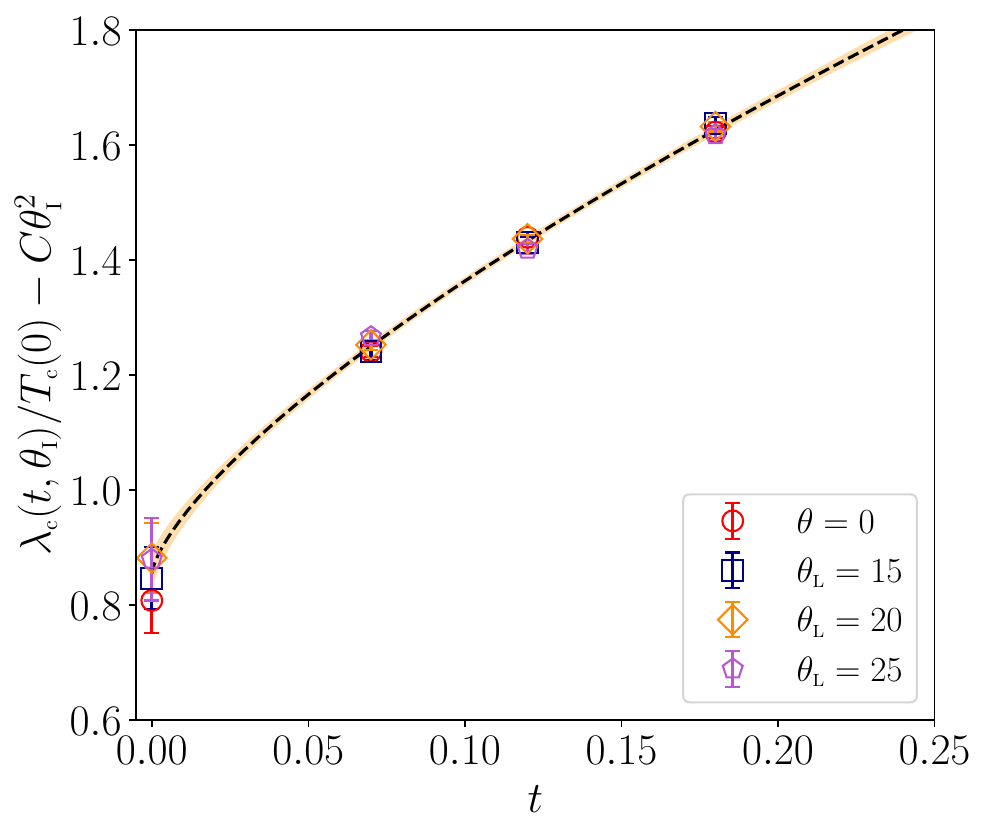}
  \caption{Global fit of the data for the mobility edge using the
    function in Eq.~\eqref{eq:global_fit_function}.  After subtracting
    the temperature-dependent term $A_1 t^{A_2}$, all data appear to
    collapse on a common curve as a function of $\theta_{\I}$ (left
    panel). Analogously, after subtracting the correction
    $C \theta_{\I}^2$ all data collapse on a common curve as a
    function of $t$ (right panel).}
  \label{fig:combo_fit}
\end{figure*}

We then proceeded to measure the mobility edge above
$T_{\c}(\theta_{\I})$ on lines of constant reduced temperature $t$,
again using the crossing point of $I_{s_0}$ with its critical value
for its determination. We will denote the mobility edge at constant
reduced temperature as
$\lambda_{\c}(t,\theta_{\I})\equiv
\lambda_{\c}|_{T=(1+t)T_{\c}(\theta_{\I}),\,\theta_{\I}}$.  Our
results for the mobility edge in lattice units, $a\lambda_{\c}$, are
reported in Tab.~\ref{tab:summary_points}, including those at $t=0$
discussed above. In Fig.~\ref{fig:mod_indep} (left panel) we show the
dimensionless ratio $ \lambda_{\c}(t,\theta_{\I})/\tcz$ against
$\theta_{\I}^2$, for all the available values of $t$. The linear trend
is clear. A quadratic dependence of $\lambda_{\c}(t,\theta_{\I})/\tcz$
on $\theta_{\I}$ is in line with the usual behavior expected of
observables at finite imaginary $\theta$-angle, under the assumption
that $CP$ is not spontaneously broken at $\theta=0$. However, this
dependence is not at all obvious, as the mobility edge is not an
observable in the conventional sense (i.e., it is not some local, or
even nonlocal functional of the gauge fields), and even under this
assumption it need not be an analytic even function of $\theta_{\I}$.

Coherently with the increase of $T_{\c}(\theta_{\I})$ and of
$L(t,\theta_{\I})$ as functions of $\theta_{\I}$, also the mobility
edge at fixed $t$ increases as a function of $\theta_{\I}$, with
approximately the same rate of increase observed for all the explored
values of $t$. For each $\theta_{\I}$ we then subtracted the value of
the mobility edge obtained at the corresponding critical temperature,
$ \lambda_{\c}^{\c}(\theta_{\I}) \equiv
\lambda_{\c}|_{T=T_{\c}(\theta_{\I}),\,\theta_{\I}}
=\lambda_{\c}(0,\theta_{\I})$, to check if the result had a simple
dependence on the reduced temperature $t$.  Subtractions were made
again between quantities measured at the same $\theta_{\L}$ rather
than at the same $\theta_{\I}$, which accounts for at least part of
the residual scatter of data points.\footnote{Note that the
  subtraction of the mobility edge at $t=0$ is mostly meant for
  illustrative purposes and to motivate the fit with
  Eq.~\eqref{eq:global_fit_function}. In this fit the systematic
  effects due to the mismatch in $\theta_{\I}$, that are at worst
  comparable with the statistical errors, are obviously
  irrelevant. Similar considerations apply for the Polyakov loop,
  discussed above in Sec.~\ref{sec:num_PL}.}  In spite of this, the
collapse of data points on a single, power-law curve as a function of
$t$ is quite clear (see Fig.~\ref{fig:mod_indep}, right panel). This
shows that the change of the mobility edge from its value at
criticality is a good, $\theta_{\I}$-independent measure of the
ordering of the system -- basically as good as the reduced
temperature.

Motivated by these findings, we performed a global fit of all our
numerical results for the mobility edge divided by $\tcz$ with an
ansatz of the general form
$ \lambda_{\c}(t,\theta_{\I})/\tcz =
\lambda_{\c}^{\c}(\theta_{\I})/\tcz + f(t)$, choosing specifically the
functional form
\begin{equation}
  \label{eq:global_fit_function}
  \f{\lambda_{\c}(t,\theta_{\I})}{\tcz}
  = \f{\lambda_{\c}^{\c}(0)}{\tcz}\left(1 + C\theta_{\I}^2\right) + A_1
  t^{A_2}\,.
\end{equation}
This choice corresponds to assuming that the mobility edge is
determined by two contributions. The first contribution comes from the
mobility edge at the $\theta_{\I}$-dependent critical temperature,
$\lambda_{\c}^{\c}(\theta_{\I})$, that is taken to depend
quadratically on $\theta_{\I}$. The other contribution comes from the
deviation from the critical line measured by the reduced temperature,
$t$, and is assumed to be a power law in $t$ vanishing at $t=0$. Our
best fit yields the following parameters,
\begin{equation}
  \label{eq:fitres}
  \begin{aligned}
    \f{\lambda_{\c}^{\c}(0)}{\tcz}
    &= 0.859(29)\,,&&& C &= 0.0608(28)\,,\\
    A_1 &= 2.61(10)\,, &&& A_2 &= 0.713(40)\,.
  \end{aligned}
\end{equation}
The best fit has a reduced chi-squared of $\simeq 0.89$ with 12
degrees of freedom, corresponding to a $p$-value of $\simeq 56\%$.
The fitted value of $\lambda_{\c}^{\c}(0)/\tcz$ is in good agreement
with its direct determination, $\lambda_{\c}^{\c}(0)/\tcz=0.808(64)$
(see Tab.~\ref{tab:summary_points}). This fit is pretty robust under
the addition of further $\mathcal{O}\left(\theta_{\I}^2\right)$
corrections to the $A_1$ and $A_2$ parameters, as such terms turn out
to be compatible with zero within errors. Moreover, their inclusion
does not change the values of $\lambda_{\c}^{\c}(0)/\tcz$, $C$, $A_1$,
and $A_2$ within errors. The results of this fit are illustrated in
Fig.~\ref{fig:combo_fit}.

\begin{figure}[t!]
  \centering
  \includegraphics[scale=0.45]{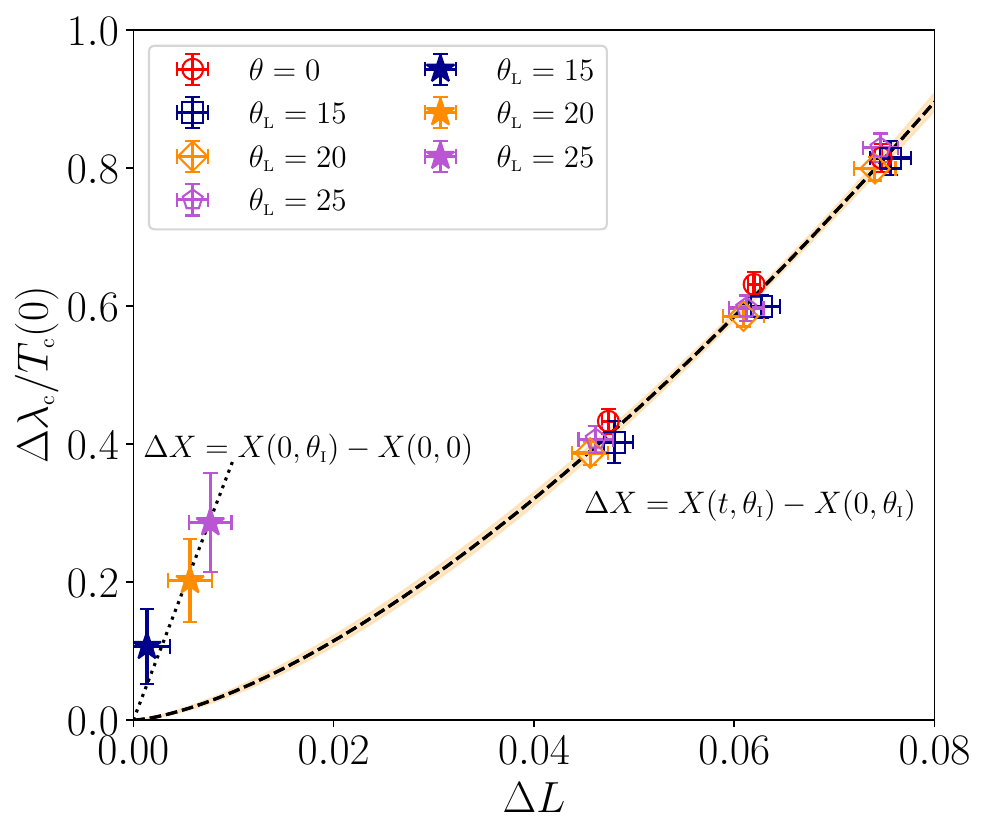}
  \caption{Correlation between the variations
    $\Delta\lambda_{\c}/\tcz$ of the mobility edge and $\Delta L$ of
    the expectation value of the modulus of the spatially averaged
    Polyakov loop [see Eq.~\eqref{eq:PLev}], relative to the critical
    line for all $t\neq 0$ (empty points), and relative to
    $\theta_{\I}=0$ along the critical line $t=0$ for all
    $\theta_{\I}\neq 0$ (filled points). The dashed and dotted lines
    show the results of a power-law and a linear fit to all the data,
    respectively [see Eqs.~\eqref{eq:lcP2} and \eqref{eq:lcP1}].}
  \label{fig:lc_pol}
\end{figure}

\begin{figure}[t!]
  \centering
  \includegraphics[scale=0.45]{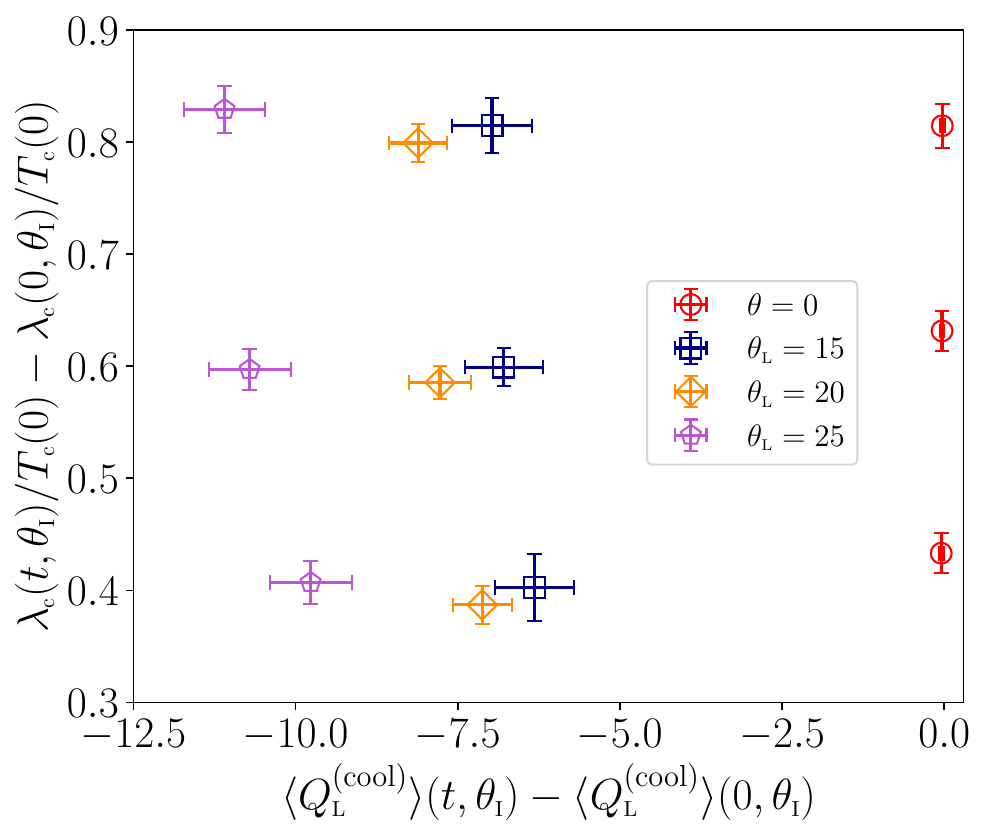}
  \caption{Relation between the mobility edge and the expectation
    value of the topological charge minus their respective values on
    the critical line.}
\label{fig:lc_q}
\end{figure}

It is worth noting that Ref.~\cite{Kovacs:2021fwq} found
$ \lambda_{\c}^{\c}(0)/\tcz
=\lambda_{\c}|_{T=\tcz,\,\theta_{\I}=0}/\tcz\sim 1$ in pure
$\mathrm{SU}(3)$ gauge theory, analyzing $N_t=8$ gauge configurations
with valence overlap fermions. Given that these results for the
mobility edge are not renormalized,\footnote{The mobility edge
  renormalizes like the quark mass, see
  Refs.~\cite{Kovacs:2012zq,Giordano:2022ghy}.} and given that they
are obtained at different lattice spacings and for two different
discretizations of the lattice Dirac operator, these two numbers do
not need to agree. Nevertheless, they fall in the same ballpark.

\subsection{Correlation of the mobility edge with the Polyakov loop
  and the topological charge}
\label{sec:num_corr}

As the mobility edge is expected to be ``dragged'' by the pseudogap
opened in the spectrum by the ordering of the Polyakov loop, it is
interesting to study the correlation between $\lambda_{\c}$ and the
expectation value of the magnitude of the spatially averaged Polyakov
loop, $\PE$. In Fig.~\ref{fig:lc_pol} we show (with empty points)
$\lambda_{\c}$ against $\PE$, after subtracting from both quantities
their value on the critical line, i.e.,
$\lambda_{\c}(t,\theta_{\I})- \lambda_{\c}(0,\theta_{\I})$ is plotted
against $\PE(t,\theta_{\I})-\PE(0,\theta_{\I})$.  (Once more,
subtractions are made between quantities measured at the same
$\theta_{\L}$ rather than at the same $\theta_{\I}$.) The strong
correlation between the two quantities is clear. Notice that the main
source of uncertainty comes from the values of both $\lambda_{\c}$ and
$\PE$ on the critical line (see Figs.~\ref{fig:pol_th}, left panel,
and \ref{fig:mod_indep}, left panel).

In contrast, no such strong correlation is found between the mobility
edge and the average topological charge. This is shown in
Fig.~\ref{fig:lc_q}, where we plot $\lambda_{\c}$ against
\begin{equation}
  \label{eq:avQ}
  \left\la Q_{\L}^{\mathrm{(cool)}}\right\ra  (t,\theta_{\I}) \equiv
  \left.\left\la Q_{\L}^{\mathrm{(cool)}}\right\ra\right|_{T=(1+t)T_{\c}(\theta_{\I}),\,\theta_{\I}}\,,
\end{equation}
again after subtracting from both quantities their value on the
critical line [of course $\la Q_{\L}^{\mathrm{(cool)}}\ra(0,0) =
0$]. Here we use $Q_{\L}^{(\mathrm{cool})}$, obtained by measuring
$Q_{\L}$ [see Eq.~\eqref{eq:top_charge}] after 30 cooling steps,
without rounding to the nearest integer.  On the contrary, the
correlation between the two quantities is rather weak at fixed $t$:
indeed, the three sets of data points lying approximately on the same
horizontal line correspond to the same $t$.

Since both $\lambda_{\c}$ and $\PE$, after subtracting their value at
criticality, grow as powers of $t$ (see Figs.~\ref{fig:pol_th} and
\ref{fig:mod_indep}), the relation between them will also be a power
law. A fit of the data in Fig.~\ref{fig:lc_pol} with a power law gives
\begin{equation}
  \label{eq:lcP2}
  \f{\lambda_{\c}(t,\theta_{\I})}{\tcz}-\f{\lambda_{\c}^{\c}(\theta_{\I})}{\tcz}
  = 
  a\left[\PE(t,\theta_{\I})-\PE(0,\theta_{\I})\right]^b  \,,
\end{equation}
with $a=37.9(5.7)$ and $b=1.482(55)$, and a reduced chi-squared of
$\simeq 7.3/10$. The exponent is in agreement with those obtained
fitting $\lambda_{\c}$ and $\PE$ against $t$ separately (see
Figs.~\ref{fig:pol_th} and \ref{fig:mod_indep}).

Along the critical line, $T_{\c}(\theta_{\I})$, the mobility edge
depends linearly on the Polyakov loop as a consequence of the fact
that both quantities are quadratic in $\theta_{\I}$ in the explored
range (see Fig.~\ref{fig:lc_pol}, filled points). This is not as
trivial as it seems, since the quadratic dependence of
$\lambda_{\c}^{\c}(\theta_{\I})/\tcz$ on $\theta_{\I}$ is not dictated
by symmetry and analyticity arguments, as pointed out above. The
relation is in this case
\begin{equation}
  \label{eq:lcP1}
  \f{\lambda_{\c}^{\c}(\theta_{\I})}{\tcz}-\f{\lambda_{\c}^{\c}(0)}{\tcz}
  = s\left[\PE(0,\theta_{\I})-\PE(0,0)\right]\,, 
\end{equation}
with $s = 37.8(6.9)$ obtained by a linear fit (see
Fig.~\ref{fig:lc_pol}).

These results are quite remarkable, and show that the mobility edge is
essentially dragged by the Polyakov loop getting more ordered, both
when moving along the critical line and when departing upward in
temperature from it. It also shows, however, that the response of the
mobility edge to Polyakov-loop ordering is different in the two cases,
with a higher sensitivity along the critical line than when moving
away from it (see Fig.~\ref{fig:lc_pol}). This likely reflects the
different response of the spectrum to the ordering of the Polyakov
loop when this is accompanied by an increase (along the critical line)
or a decrease (away from it) in the topological content of gauge
configurations -- recall that for small $\theta_{\I}$ one has
\begin{equation}
  \label{eq:Qtopcrit}
  \begin{aligned}
    \left\la Q_{\L}^{\mathrm{(cool)}}\right\ra (t,\theta_{\I})
    &    = \theta_{\I} \left.\left\la
      Q_{\L}^{\mathrm{(cool)}\,2}\right\ra\right|_{T=(1+t)T_{\c}(0),\,\theta_{\I}=0}\\
    & \phantom{=}+O(\theta_{\I}^3)  \,,
  \end{aligned}
\end{equation}
with $\la Q_{\L}^{\mathrm{(cool)}\,2}\ra|_{T,\,\theta_{\I}=0}$
decreasing with increasing $T$.  We elaborate further on this point in
the next section.  Here we want to stress that in both cases the
effects of topology on $\lambda_{\c}$ are only indirect, and mediated
by the ordering of the Polyakov loop. This is the main conclusion of
our analysis.

\section{Discussion}
\label{sec:disc}

Figures~\ref{fig:mod_indep} and \ref{fig:combo_fit} show that regarded
as a function of $\theta_{\I}$ and the reduced temperature $t$, rather
than $\theta_{\I}$ and the temperature $T$, the dimensionless mobility
edge $\lambda_{\c}(t,\theta_{\I})/\tcz $ is given by two independent
pieces,
$\lambda_{\c}(t,\theta_{\I})/\tcz =\lambda_{\c}^{\c}(\theta_{\I})/\tcz
+ f(t) $, each dependent on a single variable.  This separation into
two parts, one corresponding to the behavior of the system at
criticality for a given $\theta_{\I}$, and one depending on the
``effective distance'' of the thermal state of the system from its
critical state at that $\theta_{\I}$, as measured by
$t(T,\theta_{\I})$, reflects an analogous separation in the
expectation value of the Polyakov loop (see Fig.~\ref{fig:pol_th}). At
fixed $\theta_{\I}$, the reduced temperature, the Polyakov-loop
expectation value, and the position of the mobility edge are then all
good measures of the distance from criticality, at least in the
explored range of $\theta_{\I}$ and $t$.

The observed increase with $\theta_{\I}$ of the mobility edge at
criticality, $\lambda_{\c}^{\c}(\theta_{\I})$, follows linearly an
analogous increase in the discontinuity of the Polyakov-loop
expectation value at the transition (an increase that we demonstrate
here quantitatively for the first time, to the best of our
knowledge). Moreover, the increase of $\lambda_{\c}$ with the reduced
temperature at fixed $\theta_{\I}$, encoded in the function $f$, is
proportional to a power of the corresponding increase of the
Polyakov-loop expectation value. This shows how close the connection
between the mobility edge and the Polyakov loop is: in practice, they
provide similar measures of the ordering of the system.

The relation between topology, Polyakov-loop ordering, and behavior of
the mobility edge deserves further discussion.  First of all, we note
that topology does not directly increase the ordering of the Polyakov
loop, quite the opposite: increasing topology by increasing
$\theta_{\I}$ at fixed $T$ disorders the Polyakov loop. This results
in an increase of $T_{\c}(\theta_{\I})$ with $\theta_{\I}$, as the
transition is delayed in temperature; and in the reduction of the
effects of increasing the temperature on the thermal state of the
system, by reducing its effective distance (measured by the reduced
temperature, $t$) from the critical state.  The increase of the
expectation value of the Polyakov loop, both along the critical line
and when moving upward in temperature from it, is instead driven by
the increase in $T$.  Along the critical line the increase of the
Polyakov loop is a side effect of the delayed transition, that takes
place at a higher $T$ leading to a higher jump in its expectation
value. Moving away from criticality by increasing $t$, the Polyakov
loop obviously increases as a thermal effect, but the increase in
topology at nonzero $\theta_{\I}$ slows down its ordering with
temperature, reducing the rate of increase of the Polyakov loop with
$T$ by reducing the relevant thermal variable, $t$.

What an increase in $\theta_{\I}$ directly does along the critical
line is increase the amount of topological objects present at the
transition, an amount that is then reduced by increasing $t$. This
should explain the different sensitivity of the mobility edge to the
ordering of the Polyakov loop when moving in these two
directions. Proceeding along the critical line, the larger amount of
topological objects is expected to increase the density of near-zero
(``peak'') modes, rearranging the spectrum by removing modes from the
bulk.  This rearrangement of the spectrum is clearly seen in the data
(the behavior for $t=0$ is similar to that for $t=0.07$ shown in
Fig.~\ref{fig:spdens}). This reduces the bulk's resistance to the
Polyakov loop dragging up the mobility edge, thus making $\lambda_{\c}$
more sensitive to the Polyakov-loop ordering. Moving away from the
critical line, instead, the amount of topological objects decreases
and the density of near-zero modes with it, the bulk opposes more
resistance to the push of the Polyakov loop toward a larger mobility
edge, and the effectiveness of Polyakov-loop ordering in dragging up
the mobility edge is reduced.

It is worth noting in this context that both low-mode localization and
the appearance of a near-zero spectral peak crucially depend on the
opening of the pseudogap, and so on deconfinement.  Concerning
localization, the pseudogap provides a region of low mode density
similar to a spectrum edge, where localized modes start to appear as
soon as some disorder is present in the system. The reason is that
eigenvalues within the pseudogap correspond to wave functions
supported on rare, localized fluctuations, which do not mix easily
with modes above the pseudogap for ``energetic'' reasons, and do not
mix easily with other modes of similar eigenvalue due to the typically
large spatial separation. For the spectral peak, the opening of the
pseudogap prevents mixing of the approximate zero modes associated
with localized topological objects with non-topological modes, again
due to energetic reasons.

\section{Conclusions}
\label{sec:concl}

In this paper we have shown that the connection between deconfinement
and localization of the low-lying Dirac modes observed in a large
variety of gauge
theories~\cite{Gockeler:2001hr,Gattringer:2001ia,GarciaGarcia:2005vj,
  GarciaGarcia:2006gr,Gavai:2008xe,Kovacs:2009zj,Kovacs:2010wx,
  Kovacs:2012zq,Giordano:2013taa,Nishigaki:2013uya,Ujfalusi:2015nha,
  Cossu:2016scb,Giordano:2016nuu,Kovacs:2017uiz,Holicki:2018sms,
  Giordano:2019pvc,Vig:2020pgq,Bonati:2020lal,Baranka:2021san,
  Kovacs:2021fwq,Cardinali:2021fpu,Baranka:2022dib,Kehr:2023wrs,
  Baranka:2023ani,Bonanno:2023mzj,Baranka:2024cuf,Giordano:2021qav}
exists also in the presence of a topological term in the action. This
allowed us to study quantitatively the effects of topology on the
localization properties of the Dirac modes, by changing the amount of
topological excitations in the system.

The main conclusion of this study is that the effects of increasing
the topological content of gauge configurations by dialing up an
imaginary $\theta$-angle clearly separates into two components.  At
constant reduced temperature
$t(T,\theta_{\I})=T/T_{\c}(\theta_{\I})-1$, the effect of
$\theta_{\I}$ is to drag $\lambda_{\c}$ up in the spectrum, following
linearly the Polyakov loop as this gets more ordered. On top of this,
the mobility edge further increases with $t(T,\theta_{\I})$, again due
to the ordering of the Polyakov loop but this time with temperature,
and again following the ordering of the Polyakov loop, although with a
somewhat reduced sensitivity.  In both cases, then, the effect of
topology on the mobility edge is only indirect, and mediated by the
ordering of the Polyakov loop.

There are several lessons one can draw from this about the relation
between localization and topology, both at zero and nonzero
$\theta_{\I}$. The first one is that the appearance of a mobility edge
in the bulk of the spectrum at the deconfinement transition is very
unlikely to be caused by the corresponding change in the local
topology. This is not really surprising, as one already expects from
the sea/islands picture of localization that the appearance of this
mobility edge is mostly due to the ordering of the Polyakov loop,
which depletes the spectrum near the origin (except for the singular
peak) and makes it qualitatively similar to a spectrum edge, where
eigenmode localization is generally expected. This is also suggested
by the appearance of a mobility edge at the deconfinement transition
in gauge theories without a nontrivial topological charge, such as 2+1
dimensional pure $\mathrm{SU}(3)$ gauge
theory~\cite{Giordano:2019pvc}, or theories with discrete gauge
group~\cite{Baranka:2021san,Baranka:2022dib,Baranka:2024cuf}.

The other, more interesting lesson drawn from the influence of
$\theta_{\I}$ on the mobility edge being only indirect is that
fluctuations in the local topology are also unlikely to significantly
affect the localization properties of modes below $\lambda_{\c}$ but
well above the near-zero peak. The existence of instantons is then not
only not necessary for low-mode localization in gauge theories, as
already remarked above, but it has also little to no direct effect on
the mobility edge. The relevance of topological fluctuations for the
physics of localization near the mobility edge $\lambda_{\c}$ stems
mostly from their general disordering effect on the Polyakov loop, and
so local topological fluctuations are not expected to play any
distinguished role in supporting localized modes near
$\lambda_{\c}$. This agrees with the suggestion that these
fluctuations affect mostly the near-zero, singular-peak region of the
Dirac spectrum~\cite{Vig:2021oyt}, and with the modeling of this
region in terms of the mixing of the zero modes associated with a
dilute gas of topological objects~\cite{Edwards:1999zm,Kovacs:2017uiz,
  Vig:2021oyt,Kovacs:2023vzi}. This is also supported by our findings
concerning the enhancement of the near-zero spectral density as the
topological content of configurations is increased.

Our results justify \textit{a posteriori} the use of staggered
fermions to study the effects of topology on the near-bulk region of
the Dirac spectrum: although staggered fermions are known to have
``bad'' chiral and topological properties, meaning that they are not
ideal to detect the fine details of chiral and topological effects
unless the lattice spacing is sufficiently small,\footnote{This does
  not mean that staggered fermions are unsuitable to study the
  topological properties of gauge fields. See, e.g.,
  Ref.~\cite{Bonanno:2019xhg} for a calculation of the topological
  susceptibility in zero-temperature pure $\mathrm{SU}(3)$ gauge
  theory from the low-lying staggered spectrum.} the decoupling of
topology from the near-bulk localization physics indicates that this
is not a problem for a correct description of the behavior of the
mobility edge.

Our study provides also one more piece of evidence for the close
connection between localization and deconfinement, with a mobility
edge appearing precisely at the deconfinement temperature,
$T_{\c}(\theta_{\I})$, and confirms once more the expectations of the
sea/islands picture of localization~\cite{Bruckmann:2011cc,
  Giordano:2015vla,Giordano:2016cjs,Giordano:2016vhx,Giordano:2021qav,
  Baranka:2022dib}. In particular, the strong correlation between the
position of the mobility edge and the Polyakov-loop expectation value
provides clear evidence of the fundamental role played by the Polyakov
loop and its ordering in the localization of low Dirac modes: an
increase in Polyakov-loop expectation value widens the spectral
pseudogap, and drives the mobility edge up in the spectrum.

The natural extension of this study is to look in the place in the
spectrum where topological effects are instead expected to play a
central role, namely the near-zero region where a singular peak in the
spectral density has been observed~\cite{Edwards:1999zm,Cossu:2013uua,
  Dick:2015twa,Alexandru:2015fxa,Tomiya:2016jwr,Kovacs:2017uiz,
  Aoki:2020noz,Vig:2021oyt,Alexandru:2021pap,Alexandru:2021xoi,
  Alexandru:2019gdm,Ding:2020xlj,Kaczmarek:2021ser,Kovacs:2021fwq,
  Meng:2023nxf,Kaczmarek:2023bxb,Alexandru:2024tel}. This is left for
future work.

\begin{acknowledgments}
  We thank T.~G.~Kov{\'a}cs for numerous discussions. The work of CB
  is supported by the Spanish Research Agency (Agencia Estatal de
  Investigaci\'on) through the grant IFT Centro de Excelencia Severo
  Ochoa CEX2020-001007-S and, partially, by the grant
  PID2021-127526NB-I00, both of which are funded by
  MCIN/AEI/10.13039/501100011033.  MG was partially supported by the
  NKFIH grants K-147396 and KKP-126769, and by the NKFIH excellence
  grant TKP2021-NKTA-64. Numerical calculations have been performed on
  the \texttt{Finisterrae~III} cluster at CESGA (Centro de
  Supercomputaci\'on de Galicia).
\end{acknowledgments}

\bibliographystyle{apsrev4-2}
\bibliography{references_theta}

\end{document}